\documentclass[twoside,11pt]{article}
\usepackage{geometry}
\usepackage[utf8]{inputenc}
\usepackage{amsmath,amsfonts,amssymb,amsthm}
\usepackage[normalem]{ulem}
\usepackage{jmlr2e,tikz}

\usepackage{color}
\usepackage{subcaption}
\usepackage{setspace}
\usepackage{mathtools} 
\usepackage{scrextend}
\usepackage{diagbox}
\usepackage{hyperref}
\newcommand{\TD}{{\rm TD}}
\newcommand{\supp}{{\rm supp}}
\newcommand{\R}{{\mathbb R}}
\newcommand{\wh}{\widehat}
\renewcommand{\hat}{\wh}

\begin{document}
\title{Antimodes and Graphical Anomaly Exploration via Adaptive Depth Quantile Functions}
\author{\name Gabriel Chandler \email gabriel.chandler@pomona.edu\\
\addr Department of Mathematics and Statistics\\ Pomona
College\\ Claremont, CA 91711, USA 
\AND
\name Wolfgang Polonik
\email wpolonik@ucdavis.edu
\\ 
\addr Department of Statistics\\
University of California\\ Davis, CA 95616, USA
}
\editor{}
\maketitle 
%
\begin{abstract}%
This work proposes and investigates a novel method for anomaly detection and shows it to be competitive in a variety of Euclidean and non-Euclidean situations. It is based on an extension of the depth quantile functions (DQF) approach. The DQF approach encodes geometric information about a point cloud via functions of a single variable, whereas each observation in a data set can be associated with a single such function.  Plotting these functions provides a very beneficial visualization aspect. This technique can be applied to any data lying in a Hilbert space.

The proposed anomaly detection approach is motivated by the geometric insight of the presence of anomalies in data being tied to the existence of antimodes in the data generating distribution. Coupling this insight with novel theoretical understanding into the shape of the DQFs gives rise to the proposed adaptive DQF (aDQF) methodology. Comparisons with competitors through applications to various data sets illustrate the DQF and aDQF's strong anomaly detection performance, and the benefits of its visualization aspects.
\end{abstract}


\maketitle

\section{Introduction}
Anomaly detection (alternatively: outlier detection, novelty detection, etc.) can be regarded as the identification of rare observations in a data set that differ significantly from the remaining bulk of the data. Detection of anomalies is an important task, as anomalies can be related to problematic behavior, or to novelties, potentially driving scientific innovation, etc. There is no standard definition of an outlier or anomalous observation, though the definition provided by Hawkins (1980) captures the essence used in the current work:
%
\begin{quote}
{\em \dots an observation which deviates so much from other observations as to arouse suspicion it was generated by a different mechanism.}
\end{quote}
%
Anomalous observations can be defined in terms of either density or distance.  For instance, Hyndman (1996) and others have defined outliers as those lying in low density regions, while others (e.g. Burridge and Taylor, 2006, and Wilkinson, 2017) define them as observations lying far from the bulk of the data. 

Considering these different points of view on outliers or anomalies, one sees that rare (anomalous) events obviously correspond to low density.  Furthermore, the idea of an alternative mechanism $Q$ generating these observations relates to mixture distributions in the sense of Huber's (1964) $\epsilon-$contamination model, where the data is generated via $(1-\epsilon)P + \epsilon Q,$  $\epsilon > 0$ small and $P$ representing the non-anomalous mechanism. Thus, under this model, the presence of anomalies corresponds with the existence of antimodes, assuming $P$ and $Q$ have modes in sufficiently separated locations. Certainly, deviating substantially from the bulk of the data means that the geometry, at least ``locally" around the anomalous point or  anomalous ``micro-cluster", differs from that of non-anomalous points. 
Additionally, at least on a heuristic level, one can expect to see an antimode of the density along the line connecting such an anomaly and a non-outlying observation, and since this antimode lies between areas of higher density, its centrality can be expected to be larger than tail regions with similar density.

We explore the depth quantile function (DQF) approach put forward in Chandler and Polonik (2021) in relation to the anomaly detection task. By exploiting the multiscale geometric information contained in the DQFs, we also provide an adaptation of the DQF, termed {\em adaptive} DQF (aDQF), that is tailored to the specific task of anomaly detection and is partly motivated by an implicitly assumed low-dimensional manifold structure for the non-anomalous points.

Due to the lack of a clear definition of what constitutes a true anomaly, a graphical summary of data can provide the practitioner with valuable information about unusual observations in their data that they might investigate  further, just as a standard box plot does in one dimension. The DQF provides such a visualization of the data that is shown to be highly beneficial for the anomaly detection task. These visualizations are standard output of an associated R package.  

\begin{figure}[h!]
    \centering
    \includegraphics[width=6in,height=2.4in]{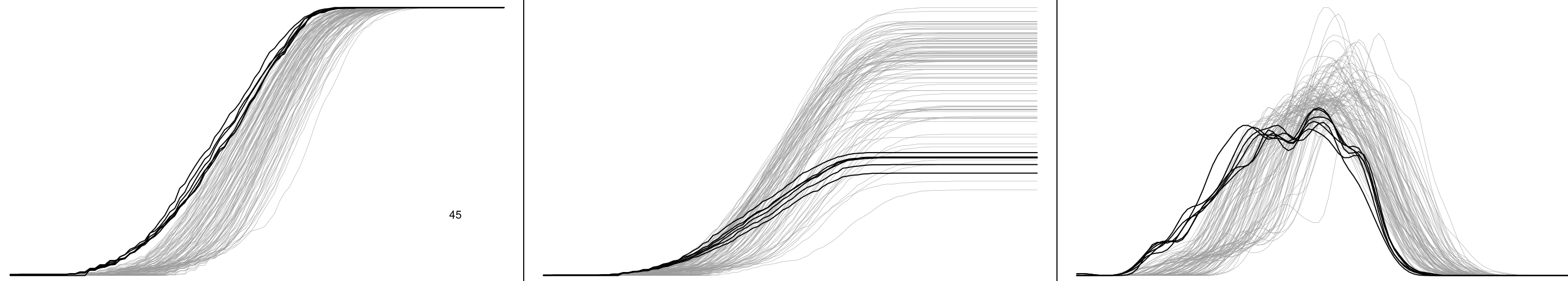}
    \caption{Functional representations via the DQF methodology for simulated data set; panels show standardized aDQFs, the aDQFs themselves, and their 1st derivatives, respectively; Data: $n=100$ observations, lying in a $d=30$ dimension annulus ($r=\frac{1}{3}, R=1$) centered at origin drawn from a density decreasing as a function of the norm. Data $x_{101},\dots,x_{105}$ form an inlying micro-cluster inside the annulus centered at (.05,\dots, .05), and $x_{106}=-x_{101}$ constitutes an isolated inlying anomaly ($d(x_{101}, x_{106})\approx 0.5$). Anomalous observations (highlighted) show different behavior than non-anomalous observations. The highlighted line isolating itself in the left most plot at large scales corresponds to the isolated anomaly. } 
    \label{fig:microcluster}
\end{figure}

It is important to note here that the visualization aspect of the DQF approach applies to any Hilbert space-valued data.  In particular, the approach applies to data lying in a Euclidean space of arbitrary dimension and, in conjunction with the kernel trick, it also applies to non-Euclidean data.  Indeed, as described below, the DQF maps Hilbert space valued data points to real valued functions of a single variable, which can then be easily plotted. Figure \ref{fig:microcluster} shows three representations of such functions, described in detail below, for a simulated data set in a 30-dimensional Euclidean space that includes various notions of anomalies (an isolated point in the ``hole'' of the annulus that forms the support of the non-anomalous observations, and a micro-cluster of five points also in the ``hole'' but away from the first anomaly). 

The relation to antimodes becomes particularly evident when considering the DQF approach for $d=1$. Therefore we present this case  which, in turn, establishes a perhaps unexpected connection to the concept of modal intervals put forward by Lientz (1974), and the related {\it shorth} plot (Sawitzki, 1994, Einmahl et al., 2010a,b).

The literature on tools for detecting anomalous observations is extensive, and it is impossible to provide a comprehensive overview. We refer to Hodge and Austin (2004),  Chandola et al. (2009), Aggarwal (2013) and Ruff et al. (2021) for surveys of both applications  of this problem and various approaches to it.  
%
%
%

Section 2 gives an introduction to the general DQF approach in both $d=1$ and higher, including discussion of tuning parameters, which yields the adaptive DQF. Section 3 considers anomaly detection in $d=1$ and elucidates the connections to existing tools, particularly the {\it shorth}.  Anomaly detection for multivariate and object data, including numerical studies on both real and simulated data, is considered in Section 4.  Section 5 presents a discussion of issues related to the implementation of this method including the accompanying R package. The final section contains technical proofs. 
\section{The DQF approach}
Suppose we observe data lying in $\R^d.$  The DQF associated with an anchor point $x \in \R^d$ is a real-valued function of a one-dimensional parameter that is generated by considering random subsets of $\R^d$ containing $x$, and computing a measure of centrality of $x$ within each random subset. This then results in a distribution of centralities (for each anchor point $x$), and the quantile function of this distribution is the DQF corresponding to the anchor point (see below for details). The anchor point can be any point that can be covered by the random subsets considered. Below we consider two types of anchor points: data points, and midpoints between pairs of data points. The latter is not only motivated by our application, but also to alleviate challenges with notions of centrality in very high dimensions. The random subsets of $\R^d$ are randomly chosen symmetric cones, and the measure of centrality used is based on Tukey's (halfspace) depth (Tukey, 1975) applied to one-dimensional projections, where for a one-dimensional distribution function $F$, Tukey depth of $x \in \R$ is defined as $TD(x,F)=\min\{F(x),1-F(x)\}$.

The essence of choosing subsets of data by only considering points captured inside a cone is related to the concept of {\em masking} that Tukey introduced with PRIM-9 (see Fisherkeller et al. 1974, Friedman and Stuetzle, 2002). The idea of masking is to systematically select subsets of the data (in other words, mask parts of the data), display low-dimensional  projections of the selected or non-masked data, and to then interactively change the mask and inspect the resulting changes in low-dimensional projections. Rather than directly visualizing projections, the DQF approach first computes a summary statistic of the data points selected by the cone. This summary statistic is the centrality of the anchor point within the selected data. The selected points (or, the complementary masked points) are altered by varying the cone and the quantile function of the summary statistic are displayed. This will be discussed further below.

\subsection{The case ${\mathbf d\;}{\boldsymbol =\;} {\mathbf 1}$}

As said above, one of the advantages of  the DQF approach is to provide a methodology for the visual exploration of geometric structure in data (including outliers, clusters, etc.) in any dimension, and even for non-Euclidean data via the kernel-trick. It turns out, however, that important insight can be gained by studying the DQF in dimension one, which also has the benefit of providing a gentle introduction to the DQF approach. 

We first introduce the so-called depth functions.  For $d=1$, the random subsets (cones) considered by the DQF approach are simply half-infinite intervals induced by a random split point $S$. For a probability measure $F$, we use the shorthand $F(x) = F((-\infty, x]), x \in \R$ to denote the distribution function. Given a point $x \in \R$, and a realization $s$ of $S$, define
\begin{align}\label{depth-funct}
d_x(s)=
\begin{cases}
\min\{F(x),F(s)-F(x)\}& \text{\rm if }\,x\leq s,\\
\min\{F(x)-F(s),1-F(x)\}& \text{\rm if }\,x\ge s,
\end{cases}
\end{align}
where we notice, for instance when $s\ge x$, $d_x(s)$ is the Tukey depth of $x$ with respect to $F$ restricted to $(-\infty,s]$. Similarly, $d_x(s)$ also is the Tukey depth of $x$ in case $s < x,$ but this time for $F$ restricted to $[s,\infty).$ So, while the notion of centrality used by the DQF approach is Tukey's (1975) data depth, it is with respect to an improper distribution, as we don't rescale the mass on the subset. 

\begin{figure}[h]
    \includegraphics[scale=0.52]{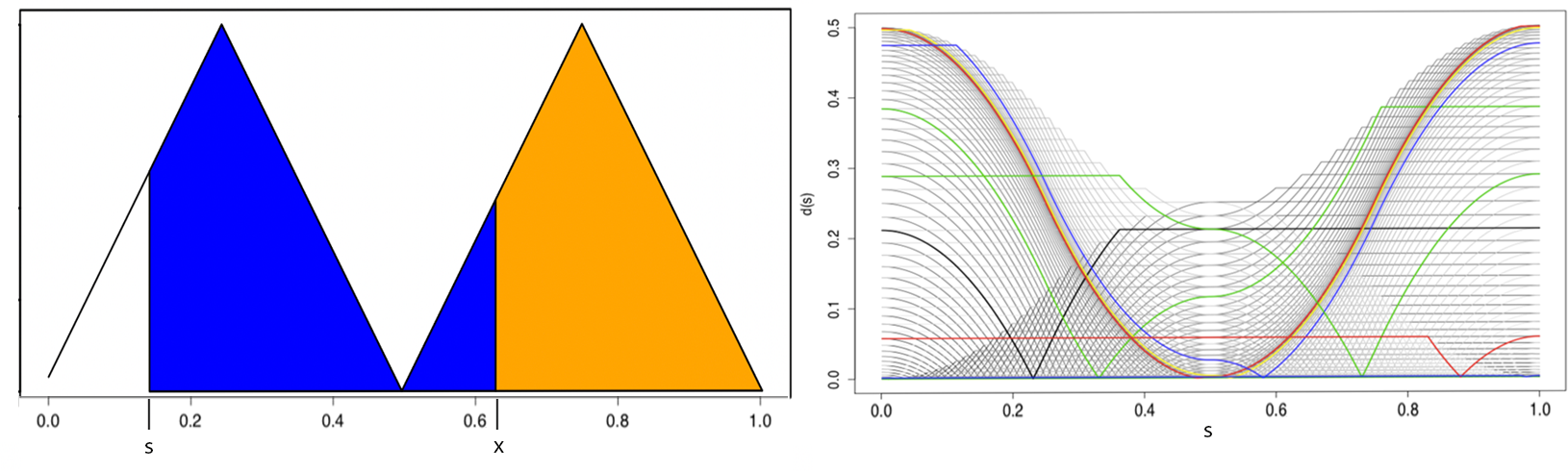}
    \caption{The function $d_x(s)$ (right) for a grid of $x$ values, based on a symmetric density $f$ comprised of two isosceles triangle (left). The associated $x$ value for each curve on the right is the point at which $d_x(s)=0$.  Colored lines added to select points $x$ to aid visualization. For the given $x$ and $s$ visualized on the left, $d_x(s)$ would be the orange shaded area $F([x,1])$ (as the smaller of the blue and the orange area). Moving $s$ further to the left would not change the value of $d_s(x)$, because though the blue area would increase, the orange would stay the same. This corresponds to flat segments seen on the left-end of the functions in the right image when $x>0.5$. }
    \label{fig:d_s}
\end{figure}

Clearly, $d_x(s)$ is non-decreasing when $s$ moves away from $x$ in either direction, and the values for $s$ on the boundaries of the support of $F$ (which could be $\pm \infty$) are both equal to the global Tukey depth  ${\rm TD}(x,F)$. Figure \ref{fig:d_s} shows functions $d_x(s)$ for a variety of $x$ values on a grid with $F$ admitting a symmetric density on $[0,1]$ comprised of two isosceles triangles.  

The random variable $S\sim G$ induces a distribution on $d_x(S)$, and the corresponding quantile function
\begin{align}
    q_x(\delta) = \inf\{t \ge 0: G(d_x(S) \le t) \ge \delta\}
\end{align}
is  the object of interest, where, for a distribution $G$ and a measurable set $A$, we let $G(A)$ denote the probability content of $A$ under $G$. The parameter $\delta$ can be thought of as measuring the scale at which information is extracted. 

For the empirical versions we use the empirical distribution $F_n$ based on the data $X_1,\ldots,X_n,$. That is, for $A \subset \R$, $F_n(A) = \frac{1}{n}\sum_{i=1}^n{\bf 1}(X_i \in A),$ i.e. $F_n(A)$ is the proportion of the data falling into $A$. Empirical depth functions $\wh d_n(s)$ and empirical DQFs $\wh q_x(\delta)$ are obtained by simply replacing $F$ with the empirical distribution $F_n$, i.e.
\begin{align}\label{emp-depth-funct}
\wh d_x(s)=
\begin{cases}
\min\{F_n(x),F_n(s)-F_n(x)\}& \text{\rm if }\,x\leq s,\\
\min\{F_n(x)-F_n(s),1-F_n(x)\}& \text{\rm if }\,x\ge s,
\end{cases}
\end{align}
where we again use the shorthand $F_n(x) = F_n((-\infty,x]),$ and 
\begin{align*}
    \wh q_x(\delta) = \inf \{t \ge 0: G(\wh d_x(S) \le t) \ge \delta\}
\end{align*}
(note that $G$ is the (marginal) distribution of $S$, so that $G(\wh d_x(S) \le t)$ depends on the data).

The following lemma addresses the {\em median property} and the {\em zero interval} property of the DQF. The median property is related to the concept of an antimode, while the presence of a zero interval is closely related to anomaly detection, as will be discussed further below. 

Let $F^\pm$ denote the left- and right-continuous generalized inverses, respectively, of the cdf $F$, i.e. $F^-(t) = \inf\{x: F(x) \ge t\}$ (left-continuous) and $F^+(t) = \inf\{x: F(x) > t\}$ (right-continuous). If $F$ has a flat part at height $t$, i.e. $F(x) = t$ for $x \in [a,b), a < b$, then $F^-(t) = a$ and $F^+(t) = b.$ 

Let $S_F \subset \R$ denote the support of $F$, and set $F^+(0) = s_*$ and $F^-(1) = s^*.$ The interval $[s_*,s^*] \cap \R$ is the smallest closed interval containing $S_F.$ 

\begin{lemma}[median property and zero interval]\label{basic} Let $F,G$ be absolutely continuous with respect to Lebesgue measure, and let the density of $G$ be strictly positive on $(s_*,s^*)$. For $ x \in S_G \cap (s_*,s^*),$ let 
\begin{align}\label{delta-star}
\delta^*_x = \begin{cases} 
G\big(F^{-}(2F(x))\big) - G(s_*), & \text{for }\;F(x) \le \frac{1}{2}\\
G(s^*) - G\big(F^+(2F(x) - 1)\big), & \text{for }\;F(x) \ge \frac{1}{2}.
\end{cases}
\end{align}
Then, for $0 \le \delta \le \delta^*_x$, there exists an interval $I_{x,\delta} = [a_{x,\delta},b_{x,\delta}] \subset S_f$ of $G$-measure $\delta$ satisfying the `median property':
\begin{align}\label{DQF-rep}
q_x(\delta)&=\frac{1}{2}F(I_{x,\delta}) = F([a_{x,\delta},x]) = F([x,b_{x,\delta}]).
\end{align} 
Moreover, 
\begin{align} \label{small-scale_large_scale}
q_x(\delta) = \begin{cases} 0 & \text{for }\;0 \le \delta \le l_x\\
\TD(x,F) & \text{for }\;\delta_x^* \le \delta \le 1\end{cases},
\end{align}
where $l_x = \sup\{G(b)-G(a):\, x \in [a,b],\,F([a,b]) = 0\}$  is the length of the `zero-interval'. 
\end{lemma}

\begin{figure}[h]

\begin{minipage}{.5\linewidth}
\centering
\subfloat[]{\label{main:a}\includegraphics[width = 2.5in,height=1.5in]{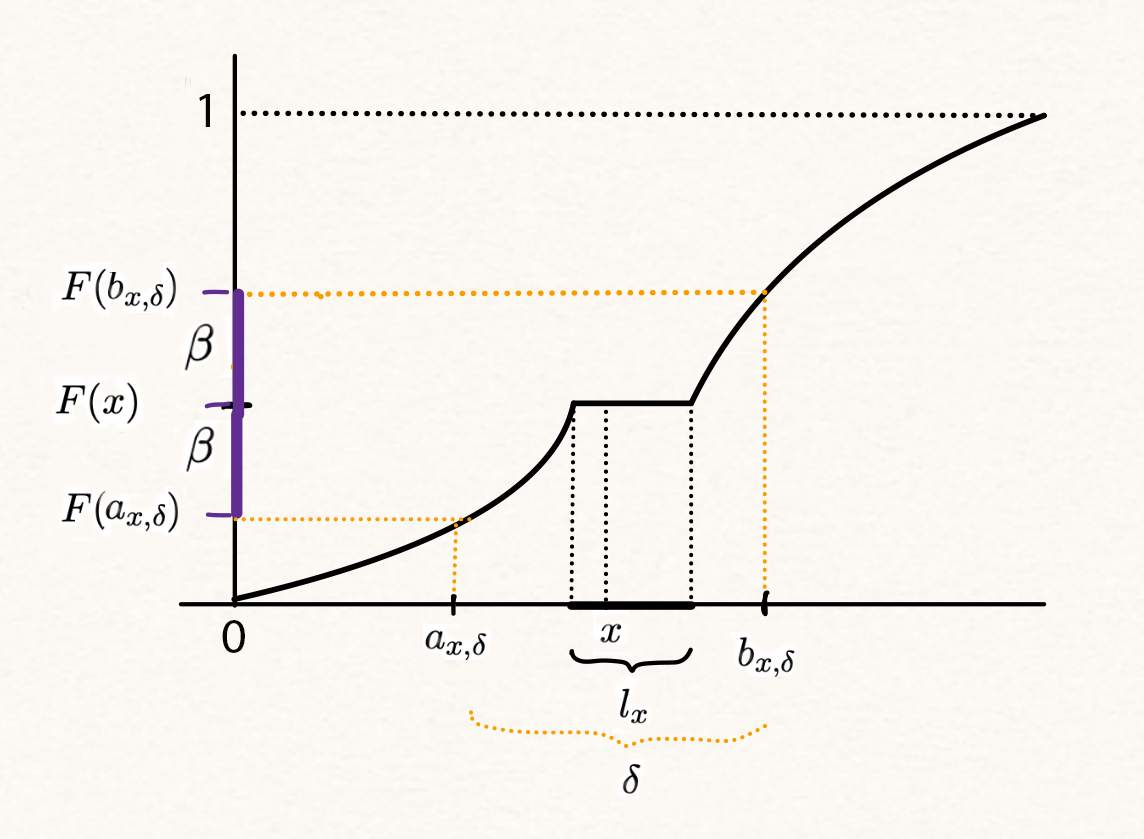}}
\end{minipage}%
\begin{minipage}{.5\linewidth}
\centering
\subfloat[]{\label{main:b}\includegraphics[width = 2.5in,height=1.5in]{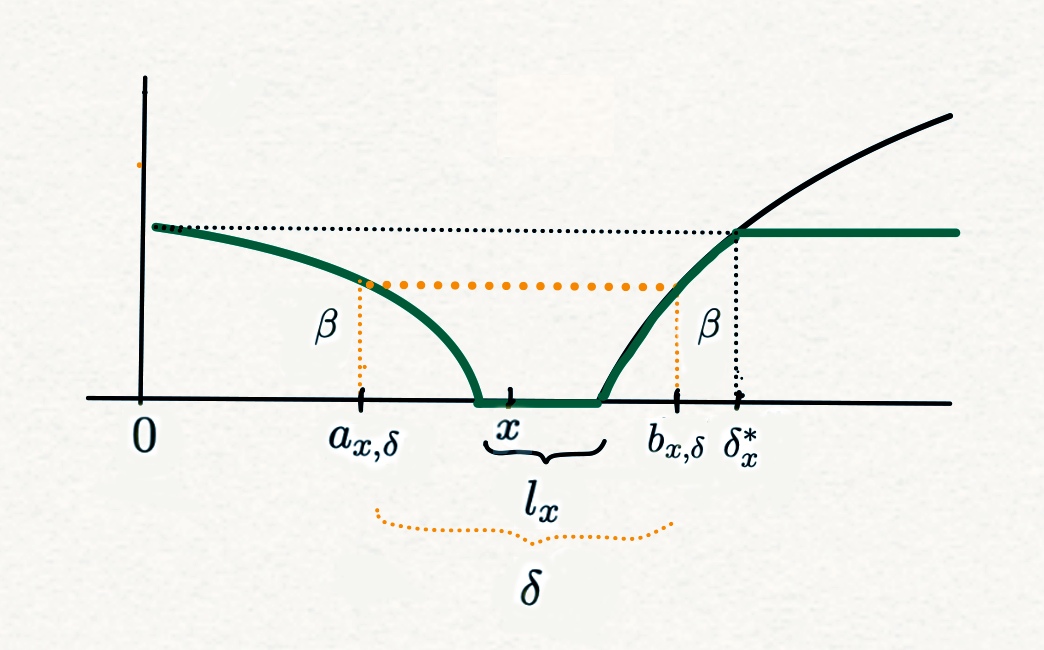}}
\end{minipage}\par\medskip
\centering
\subfloat[]{\label{main:c}\includegraphics[width = 2.5in,height=1.5in]{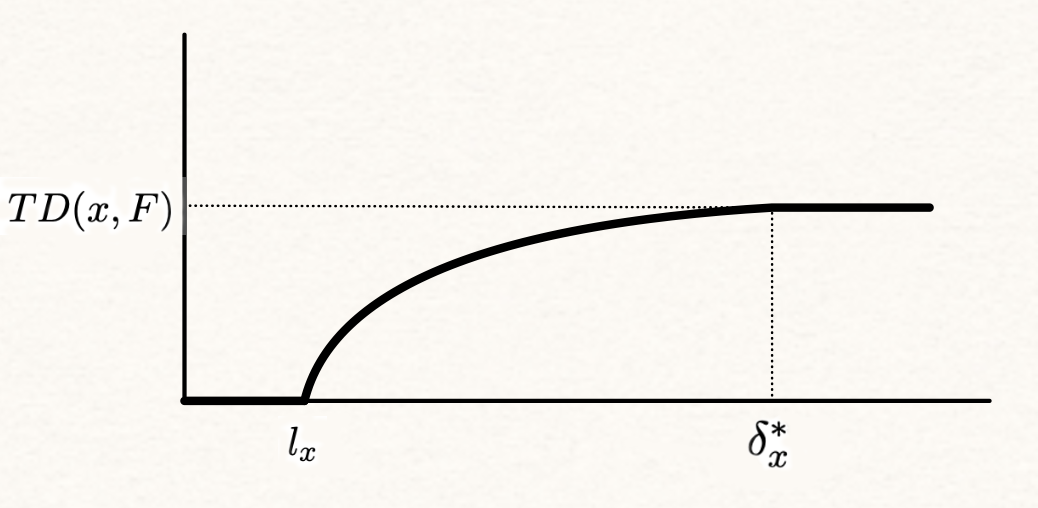}}

\caption{Illustration of Lemma~\ref{basic} with $G = U[0,1]$: Panel (a) shows the cdf $F$ with a flat part around $x$, i.e. $x$ lies outside the support; here, the length of the flat part is $l_x$; for a given $\delta > 0$, find an interval on the vertical axis symmetric about $F(x)$ (of length $2\beta$), such that its pre-image is of length $\delta;$ this pre-image is the interval $[a_{x,\delta},b_{x,\delta}]$; note that the $F$-measure of this interval by construction is $2\beta$, and having $F(x)$ in the center of the vertical interval corresponds to the median property; Panel (b) shows the corresponding depth function $d_x(s)$ (in green); as $F(x) > 1/2$, it is obtained by reflecting the cdf to the left of $x$ upwards about the horizontal axis at height $F(x)$, then shifting the graph down by the amount $F(x)$ and capping the resulting function (on the right) at height $F(x)$; the sublevel set of $d_x(s)$ at height $\beta$ now is the interval $[a_{x,\delta},b_{x,\delta}]$;  note that there is no symmetric interval with nonnegative endpoints about $F(x)$ for $\beta > F(x)$, and this leads to the flat part on the right of $d_x(\delta);$ Panel (c) shows the DQF $q_x(\delta)$, which is $\beta$ as a function of $\delta.$ }
\label{fig:geom-illustr}
\end{figure}

Figure~\ref{fig:geom-illustr} illustrates the assertions of the lemma geometrically for the case $G = U[0,1]$. In brief, without any other flat parts of $F$ except for a flat part containing $x$ (that is, a zero-interval), all the intervals $[a_{x,\delta},b_{x,\delta}]$ can be constructed either as sublevel sets of $d_x(s)$ (see Figure~\ref{fig:geom-illustr}, panel (b)), or, alternatively, they can be obtained as pre-images under $F$ of intervals with midpoint $F(x)$ grown until the $G$-measure (length, for $G=U[0,1])$) of the pre-image equals $\delta$ (see Figure~\ref{fig:geom-illustr}, panel (a)). Note that the asserted median property immediately follows from $F(x)$ being the midpoint of the image of $[a_{x,\delta},b_{x,\delta}]$ under $F$. The value  $\delta_x^*$ is the $G$-measure of the largest of the intervals $[a_{x,\delta}, b_{x,\delta}]$, and this largest interval is of the form implied by (\ref{delta-star}), that is, $[s_*, F^-(2F(x))]$ for $F(x) \le 1/2$ and $[F^+(2F(x) - 1),s^*]$ for $F(x) \ge 1/2.$ The zero-interval exists if $l_x > 0$ with $l_x$ from (\ref{small-scale_large_scale}). The zero-interval property has an analog in the multivariate case (see Lemma~\ref{DQF-atzero2}). 

The zero-interval is important in the context of anomaly detection, as it provides an explanation for the empirical observation that the behavior of the DQF for small values of $\delta$ has a discriminatory effect, and we use the length of the zero interval (or, an approximation of this) to construct an anomaly score that is used in the numerical comparisons. One of the situations in which this is of particular interest is when the anchor point $x$ is chosen as the midpoint between two observations, which might fall outside the support of $F$. Though this length serves as a simple, one-dimensional numerical feature extracted from the DQFs  that can serve as an anomaly score, we would like to stress that as a function, the DQF contains much more information. It might therefore be beneficial to also consider other values of $\delta$, or, more generally, other features of the DQFs. All of this will be discussed in more detail below.

%

\begin{corollary}\label{DQF-atzero}
Consider the setup of Lemma~\ref{basic}, and suppose that $F$ has density $F^\prime = f$. Then, for $x \in S_g$,
\begin{align*}\lim_{\delta\to 0} \frac{2q_x(\delta)}{\delta}&=\frac{f(x)}{g(x)}.
\end{align*}
\end{corollary}
The above results, more precisely, (\ref{small-scale_large_scale}) and Corollary~\ref{DQF-atzero}, show that the DQF is informative at both large and small scales, i.e. for large and small values of $\delta$, motivating the usefulness of the entire function in examining a data set. Analogous properties for the case $d \ge 2$ can be found in Chandler and Polonik (2021), Lemma 2.2 a-b). This property of being informative on all scales is in contrast to other multi-scale methods, such as the mode tree (Minnotte and Scott, 1993), which transitions from $n$ modes to 1 (thus no information on either extreme), or a multi-scale version of the shorth which we discuss and explore in section 3.


\subsection{The case ${\mathbf d\;}  {\boldsymbol >\;} {\mathbf 1}$} As localization is an important aspect of the DQF for small values of $\delta$, one finds that half-spaces as our random subsets will not suffice.  Rather, we consider symmetric cones with opening angle $\alpha$ strictly less than $\pi/2$ (the angle between any non-zero vector on the surface of the cone and the axis of symmetry). We use symmetric cones to not privilege any particular direction in our space beyond the specified axis of symmetry for the cone. The question is then how to put a distribution over cones in order to compute quantile functions.  

First we discuss the construction of the empirical version of the DQF.  For any two points in the data set, say $x_i$ and $x_j$, $i\neq j$, let $\ell_{ij}$ denote the line passing through them, and  consider (one-sided) symmetric cones $C_{ij}(s)$ with $\ell_{ij}$ as axis of symmetry, $s$ as the tip located on $\ell_{ij}$ (see Figure~\ref{dqf-1}). The fixed opening angle $\alpha$ is a tuning parameter. Formally, given $x_i,x_j$, let $s = x_j + \gamma(x_i - x_j)$ for some $\gamma \in \R.$ Then, we have $C_{ij}(s) = \big\{x\in \R^d: \frac{\langle x_i - x_j, x-s\rangle}{\|x_i-x_j\|\,\|x-s\|} \le \cos \alpha\big\}.$ To change the orientation of the cone, replace $x_i - x_j$ by $x_j - x_i$ in this definition. Select a point $m_{ij}$ on the line $\ell_{ij}$ as the ``anchor point.'' The choice of $m_{ij}$ might depend on the specific application considered. In our applications, we choose    $m_{ij}=\frac{x_i+x_j}{2}$, the midpoint, but it could also be $x_i$, or $x_j$, as we are using in the one-dimensional case discussed above. See section 4.1 for discussion of this choice. The orientation of the cone $C_{ij}(s)$ is such that $m_{ij}$ is contained inside. In other words, the orientation of the cone flips depending on which side of $m_{ij}$ the tip lies. 

\tikzset{every picture/.style={line width=0.75pt}} 
\begin{figure}
\centering
\resizebox{4in}{!}{

\begin{tikzpicture}[x=0.75pt,y=0.75pt,yscale=-1,xscale=1]

\draw   (2,1) -- (520,1) -- (520,294) -- (2,294) -- cycle ;
\draw  [fill={rgb, 255:red, 155; green, 155; blue, 155 }  ,fill opacity=0.36 ][line width=1.5]  (95.79,229.69) -- (228,139) -- (255.88,238.38) -- cycle ;
\draw [line width=1.5]    (228,139) -- (352,59) ;
\draw [line width=1.5]    (255.88,238.38) -- (396,246) ;
\draw [line width=1.5]  [dash pattern={on 1.69pt off 2.76pt}]  (352,59) -- (397,30) ;
\draw [line width=1.5]  [dash pattern={on 1.69pt off 2.76pt}]  (396,246) -- (457,249) ;
\draw  [dash pattern={on 0.84pt off 2.51pt}]  (20,249) -- (492,123) ;
\draw  [dash pattern={on 4.5pt off 4.5pt}]  (33,246) -- (375,14) ;
\draw  [dash pattern={on 4.5pt off 4.5pt}]  (33,246) -- (450,273) ;
\draw [line width=1.5]  [dash pattern={on 1.69pt off 2.76pt}]  (223,118) -- (228,139) ;
\draw [line width=1.5]  [dash pattern={on 1.69pt off 2.76pt}]  (255.88,238.38) -- (261,261) ;

\draw (149,201) node [anchor=north west][inner sep=0.75pt]   [align=left] {x};
\draw (171,188) node [anchor=north west][inner sep=0.75pt]   [align=left] {x};
\draw (170,207) node [anchor=north west][inner sep=0.75pt]   [align=left] {x};
\draw (207,160) node [anchor=north west][inner sep=0.75pt]   [align=left] {x};
\draw (222,172) node [anchor=north west][inner sep=0.75pt]   [align=left] {x};
\draw (381,173) node [anchor=north west][inner sep=0.75pt]   [align=left] {x};
\draw (283,116) node [anchor=north west][inner sep=0.75pt]   [align=left] {x};
\draw (186,77) node [anchor=north west][inner sep=0.75pt]   [align=left] {x};
\draw (363,76) node [anchor=north west][inner sep=0.75pt]   [align=left] {x};
\draw (341,185) node [anchor=north west][inner sep=0.75pt]   [align=left] {x};
\draw (427,197) node [anchor=north west][inner sep=0.75pt]   [align=left] {x};
\draw (381,79) node [anchor=north west][inner sep=0.75pt]   [align=left] {x};
\draw (451,160) node [anchor=north west][inner sep=0.75pt]   [align=left] {x};
\draw (374,229) node [anchor=north west][inner sep=0.75pt]   [align=left] {x};
\draw (150,150) node [anchor=north west][inner sep=0.75pt]   [align=left] {x};
\draw (95,210) node [anchor=north west][inner sep=0.75pt]   [align=left] {x};
\draw (200,162) node [anchor=north west][inner sep=0.75pt]   [align=left] {x};
\draw (440,112) node [anchor=north west][inner sep=0.75pt]   [align=left] {$\displaystyle \ell _{i}{}_{j}$};
\draw (16,225) node [anchor=north west][inner sep=0.75pt]   [align=left] {$\displaystyle x_{i}$};
\draw (386,126) node [anchor=north west][inner sep=0.75pt]   [align=left] {$\displaystyle x_{j}$};
\draw (12,246) node [anchor=north west][inner sep=0.75pt]    {$\bullet $};
\draw (386,146.4) node [anchor=north west][inner sep=0.75pt]    {$\bullet $};
\draw (93,233) node [anchor=north west][inner sep=0.75pt]    {$\mathbf s$};
\draw (250,190.4) node [anchor=north west][inner sep=0.75pt]    {$m_{i}{}_{j}$};
\draw (237,187.4) node [anchor=north west][inner sep=0.75pt]    {$\bullet $};

\draw (197,208.4) node [anchor=north west][inner sep=0.75pt]    {$A_{i}{}_{j}( s)$};
\draw (323,120.4) node [anchor=north west][inner sep=0.75pt]    {$B_{i}{}_{j}( s)$};
\draw (300,68) node [anchor=north west][inner sep=0.75pt]   [align=left] {x};
\draw (165,165) node [anchor=north west][inner sep=0.75pt]   [align=left] {x};
\draw (184,97) node [anchor=north west][inner sep=0.75pt]   [align=left] {x};
\draw (201,243) node [anchor=north west][inner sep=0.75pt]   [align=left] {x};
\draw (201,84) node [anchor=north west][inner sep=0.75pt]   [align=left] {x};
\draw (84,256) node [anchor=north west][inner sep=0.75pt]   [align=left] {x};
\draw (210,262) node [anchor=north west][inner sep=0.75pt]   [align=left] {x};
\draw (234,261) node [anchor=north west][inner sep=0.75pt]   [align=left] {x};
\draw (203,115) node [anchor=north west][inner sep=0.75pt]   [align=left] {x};
\draw (28,250.4) node [anchor=north west][inner sep=0.75pt]    {$\mathbf s'$};
\draw (325,13) node [anchor=north west][inner sep=0.75pt]   [align=left] {x};
\end{tikzpicture}
}
\caption{A cone $C_{ij}(s)$ in $\R^2$ with axis of symmetry $\ell_{ij}$ passing through points $x_i,x_j$ split into $A_{ij}(s)$ and $B_{ij}(s)$ with the split determined by the anchor point $m_{ij}.$ The shaded triangle (cone) is the set $A_{ij}(s)$, while $B_{ij}(s)$ is the remaining part of $C_{ij}(s)$. Only points inside $C_{ij}(s)$ are considered. Points currently outside $C_{ij}(s)$ will be considered for cone tips $s'$ further away from $m_{ij}$ as indicated. Here, $\wh d_{ij}(s) = F_n(A_{ij}(s))$ because $A_{ij}(s)$ contains fewer points than $B_{ij}(s).$}
\label{dqf-1}
\end{figure}

Given any such cone $C_{ij}(s)$,  split the cone into two parts $A_{ij}(s)$ and $B_{ij}(s) = C_{ij}(s) \setminus A_{ij}(s)$ with $A_{ij}(s)$ being the ``top part'' of the cone, obtained by cutting $C_{ij}(s)$ along the hyperplane orthogonal to $\ell_{ij}$ passing through $m_{ij}$. We sometimes also call these two type of sets $A$-sets (cones themselves) and $B$-sets (frustums). By convention, we consider the boundary between $A_{ij}(s)$ and $B_{ij}(s)$ to be part of $A_{ij}(s)$ but not of $B_{ij}(s)$ in order to make the two sets disjoint. Then, the measure of centrality of $m_{ij}$ that we are using is the depth $$\wh{d}_{ij}(s) = \min\big(F_n(A_{ij}(s)), F_n(B_{ij}(s))\big),$$
where, as in the one-dimensional case, $F_n$ denotes the empirical measure, such that $F_n(A)$ is the proportion of the data falling into $A \subset \R^d$. Equivalently, one can think of constructing $\hat d_{ij}(s)$ by projecting the data lying inside the cone $C_{ij}(s)$ onto $\ell_{ij}.$ Then $\hat d_{ij}(s)$ is the one dimensional Tukey depth of $m_{ij}$ among these projections, again with respect to the improper empirical (sub)distribution. 

Finally, we choose the tip randomly according to a distribution $G_{ij}(s)$ along the axis of symmetry $\ell_{ij}$. (Choosing $G_{ij}$ is discussed in \ref{adaptive}.) The DQF $\hat q_{ij}(\delta)$ is then defined  as the quantile function of the distribution of the random depths with respect to $G_{ij}(s)$, i.e.
$$\hat q_{ij}(\delta) = \inf\big\{t > 0:\, G_{ij}(s: \hat d_{ij}(s) \le t) \ge \delta\big\}, $$
where, for a measurable set $D \subset \ell_{ij}$ we let $G(D)$ denote the $G$-measure of $D$. Note that $G_{ij}(s: \hat d_{ij}(s) \le t)$ depends on all of the data. One might use different approaches to reduce the total number of functions $\hat q_{ij}(\delta)$ to consider. A natural approach is to use averaging: For each data point $x_i,$ average the functions $\hat{q}_{ij}(\delta)$ over $j$ (note $\hat{q}_{ii}$ does not exist), i.e. for each $x_i$, we consider
$$\overline q_{i}(\delta) = \frac{1}{n-1}\sum_{j=1 \atop j \ne i}^n \hat q_{ij}(\delta). $$
More robust approaches include winsorized averages, truncated averages, etc. A slightly different summary is to consider normalized averages of the DQFs of the form
$$\widetilde q_i(\delta) = \frac{\bar{q}_i(\delta)}{\bar{q}_i(1)}.$$
%
These normalized averages embed the global behavior of the DQF at all $\delta$ values and focus solely on the shape of the DQF.  

\begin{figure}[h]
    \centering
    \includegraphics[scale=0.5]{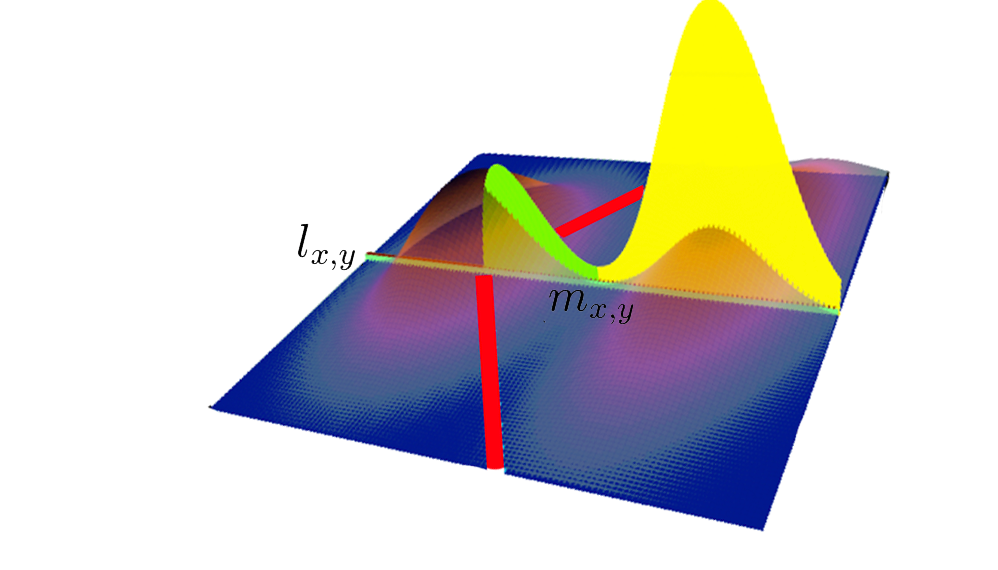}
    \caption{A 2-dimensional bi-modal density (blue to orange, transparent to allow for visualization of the lower dimensional features), a given anchor point $m_{x,y}$ (green/yellow region intersection), direction $\ell_{x,y}$ (green line in plane), and cone tip $s$ (convergence of red cone boundaries). Green/yellow function over $\ell_{x,y}$ indicates the probability mass inside the cone projected onto $\ell_{x,y}$. For this cone tip, $d_{x,y}(s)$ would equal the green area under the one-dimensional curve rather than the larger yellow area.} \label{fig:depth_pop_2d}
\end{figure}

{\sc Population versions} of the empirical depth functions $\hat d_{ij}(s)$ and the corresponding empirical DQFs are defined as follows. For $x,y \in \R^d$, let $\ell_{x,y}$ denote the line passing through both $x$ and $y$. For $s \in \ell_{x,y}$ let $C_{x,y}(s)$ denote the symmetric cone in $\R^d$ with tip $s$ containing the anchor point, say $m_{x,y} = \frac{x+y}{2},$ with axis of symmetry given by $\ell_{x,y}$, and with opening angle given by the tuning parameter $\alpha$. The hyperplane through $m_{x,y}$ perpendicular to $u$ again divides $C_{x,y}(s)$ into two sets, a cone $A_{x,y}(s)$ and a frustum $B_{x,y}(s),$ where for definiteness, the intersection of the hyperplane and the cone (the base of $A_{x,y}(s)$ or the top of $B_{x,y}(s)$) is considered to be part of $A_{x,y}(s)$ but not of $B_{x,y}(s)$. Then, 
$$d_{x,y}(s) = \min\big(F(A_{x,y}),\,F(B_{x,y})\big).$$
Similar to the empirical version, $d_{x,y}(s)$ is the one-dimensional Tukey depth of the (sub)distribution obtained by projecting the mass of the cone $C_{x,y}(s)$ onto the axis of symmetry $\ell_{x,y}$. Figure \ref{fig:depth_pop_2d} illustrates this for a two-dimensional bi-modal density. 

Again, choosing the tip randomly on the line $\ell_{x,y}$ according to $S \sim G = G_{x,y}$ results in a random variable $d_{x,y}(S)$ whose quantile function is the DQF $q_{x,y}(\delta)$. Formally,
$$q_{x,y}(\delta) = \inf \{t > 0:\, G_{x,y}( d_{x,y}(S) \le t) \ge \delta\}.$$
The empirical average $\overline q_{i}(\delta)$ corresponds to 
$$\overline q_{x}(\delta) = {\rm E}\big[q_{x,Y}(\delta)\big],\quad Y\sim F,$$
and similarly $\tilde q_{x}(\delta) = \frac{\overline q_{x}(\delta)}{\overline q_{x}(1)}$ is the normalized expected DQF. As already mentioned, a result for $d \ge 2$, analogous to Corollary~\ref{DQF-atzero} can be found in Chandler and Polonik (2021). The flatness of the DQF for anchor points lying outside the support of $F$, i.e. a result analogous to the last assertion of  Lemma~\ref{basic}, also holds for $d \ge 2,$ as shown in Lemma~\ref{DQF-atzero2}. 

We use the following setting: For a given distribution $F$ on $\R^d$ with compact support (for simplicity) and $x,y \in \R^d$, let $[a_{x,y},b_{x,y}] \subseteq \ell_{x,y}$ denote the range of the push-forward distribution on $\ell_{x,y}$ obtained by the projection map $\pi_{x,y}(z)$, which is the orthogonal projection of $z\in \R^d$ onto $\ell_{x,y}.$ We assume $\|b_{x,y} - a_{x,y}\| > 0$. Let $G_{x,y}$ denote a continuous distribution on $[a_{x,y},b_{x,y}]$. 
This is a distribution on the ``one-dimensional interval'' of length $\|b_{x,y} - a_{x,y}\|$. Also, we represent $\ell_{x,y}$ as $\ell_{x,y} = \{z\in \R^d:\, z = m_{x,y} + t u_{x,y}, t \in \R\}$, where $m_{x,y}$ is the anchor point, and $u_{x,y} \in S^{d-1}$ is the unit vector giving the direction of $\ell_{x,y}.$

\begin{lemma}[zero interval property in multivariate settings] \label{DQF-atzero2} Consider the above setting for given $x,y \in \R^d.$ Set 
\begin{align*}
t^\pm_{x,y} &= \inf_{t \ge 0}\{s = m_{x,y} \pm tu_{x,y},\; s \in [a_{x,y},b_{x,y}]\;\text{\rm and}\;F(A_{x,y}(s)) \ne 0\}\\
\shortintertext{and} 
s^\pm_{x,y} &=  m_{x,y} \pm t^\pm_{x,y}u.
\end{align*}
 Then,\\[-10pt]
$$q_{x,y}(\delta) =0\qquad\text{for }\;\; \delta \in [0,\delta_{x,y}],$$
where $\delta_{x,y} = G_{x,y}([s^-_{x,y},s^+_{x,y}]).$ In case $G_{x,y} = U[a_{x,y},b_{x,y}]$, the uniform distribution on $[a_{x,y},b_{x,y}],$ we can write
$ G_{x,y}([s^-_{x,y},s^+_{x,y}]) = \frac{t^+_{x,y} + t^-_{x,y}}{\|b_{x,y} - a_{x,y}\|}.$ 
Furthermore, we have $$\overline{q}_x(\delta) = 0 \qquad\text{for }\;\;\delta \in [0,\inf_{y}\delta_{x,y}].$$
%
%
%
\end{lemma}

\begin{figure}[h]
    \centering
    \includegraphics{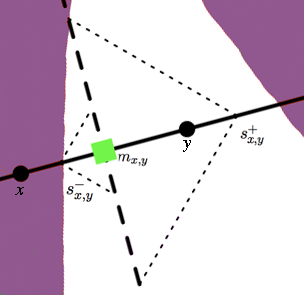}
    \caption{The point $y$ corresponds to an anomalous observation living outside the support (purple shaded region).  This point is compared to the non-anomalous $x$, and the midpoint is used as the anchor point. The length of the zero-interval, $\delta_{x,y}$,  is the $G-$measure of the interval $(s_{x,y}^-, s_{x,y}^+)$.}
    \label{fig:zeroint}
\end{figure}

Geometrically, the quantity $t^+_{x,y} + t^-_{x,y} = \|s^+_{x,y} - s^-_{x,y}\|$ is the sum of the heights of two cones $A_{x,y}(s^+_{x,y}), A_{x,y}(s^-_{x,y})$, see Figure \ref{fig:zeroint}. Note that the two cones point in opposite directions, and by definition of $t^+_{x,y}$ and $t^-_{x,y}$, they are chosen such that they give the largest symmetric cones under consideration that do not intersect with the support of $F.$ Also note that the two cones belong to the same anchor point that is lying between the bases of these two cones. 

Suppose that the support of $F$ has a hole and the anchor point $m_{x,y}$ lies in this hole (think of $m_{x,y}$ as the midpoint between $x$ and $y$), then the $G_{x,y}$-measure of the combined heights, $\delta_{x,y} = G_{x,y}([s^-_{x,y}, s^+_{x,y}]),$ measures some aspects of the geometry of the hole. Moreover, if, for a given $x$, there is a high probability that a randomly chosen $y$ is such that the corresponding anchor point lies off the support of $F$, then $\overline q_x(\delta)$ will also be small for a range of small values of $\delta.$ (Recall that everything depends on the opening angle of the cones as well.)

Also note that Lemma~\ref{DQF-atzero2} is a generalization of the one-dimensional case presented in the last part of  Lemma~\ref{basic}. In this one-dimensional case there is of course no need to consider pairs since there is only one axis. For further discussion of the population version and theoretical properties of the corresponding estimators, see Chandler and Polonik (2021). 

\subsubsection{Discussion of the (near) zero-interval property}
As indicated above, the behavior of the (average) DQF for small values of $\delta$ is quite useful for visual anomaly inspection. Indeed, in our experiments we have observed two types of anomalies that are expressed by opposite behaviors of the zero intervals: One type of anomaly leads to a longer zero interval (e.g. see Figure~\ref{fig:fnanom}), the other one to a shorter interval (see Figures~\ref{fig:microcluster}). In either case, the zero interval property is an important visual indicator of possible anomalies. The two types of behavior can be explained geometrically as is discussed in the following in the population setting. A similar interpretation holds in the sample setting.

Type 1: Anomalies that give rise to {\em longer} (near) zero intervals. Assume again an $\epsilon$-contamination model $(1-\epsilon)P + \epsilon Q,$  $\epsilon > 0$ (small), and let $S_P$ and $S_Q$ denote the supports of $P$ and $Q$, respectively. Assume that $S_P$ is convex, and $S_Q \cap S_P = \emptyset$. For simplicity, assume that $P$ is the uniform distribution on $S_P,$ and in the construction of the DQFs choose the midpoint $m_{x,y}$ as a base point. Then, by convexity, for all pairs $(x,y) \in S_P \times S_p$, the base point $m_{x,y}$ lies in $S_P,$  so there is no zero interval for such $(x,y)$. On the other hand, for pairs $(x,y) \in S_Q \times S_P$ with $m_{x,y} = (x + y)/2 \notin S_P \cup S_Q$, we do observe a zero interval. As a result,  for $x \in S_Q$ (anomalous), the average DQFs (which are used in all the numerical experiments in this work) will display smaller DQF values for $\delta$ close to zero than for $x \in S_P$ (non-anomalous). The visual power of this approach depends on how much the lengths of the (near) zero intervals differ for anomalous points $x$ versus non-anomalous points. In particular, it depends on the $G_{x,y}$-measure, and it also  depends on the geometry of $S_P$ (in this discussion, we assume $S_Q$ to be negligibly small). The adaptive DQF methodology (see below) attempts to find $G_{x,y}$ adaptively to enhance the differences in the lengths of the (near) zero interval.

Type 2: Anomalies that give rise to {\em shorter} (near) zero intervals. A constellation that gives rise to such behavior of the DQFs is the situation illustrated in Figure~\ref{fig:microcluster}. There, considering again the $\epsilon$-contamination model, the support $S_P$ is not convex. Indeed, in the example presented in  Figure~\ref{fig:microcluster}, the support $S_P$ is an annulus, and $S_Q$ lies inside the empty region. For simplicity, assume that $S_Q$ itself is also a ball with midpoint zero, but with a very small (negligible) radius. In this case, the maximum zero interval of $q_{x,y}$ with $(x, y) \in S_Q \times S_P$ is at most the $G_{x,y}$-measure of the line segment obtained by intersection $\ell_{x,y}$ with the empty ball (the actual measure depends on the opening angle of the cones used in the construction of the DQFs). In contrast to that, for many pairs $(x,y) \in S_P \times S_P,$ the zero (or near zero) interval are longer. In the extreme case, they can be as long as the $G_{x,y}$-measure of the entire line segment obtained by intersection $\ell_{x,y}$ with the empty ball. This implies that the average DQFs display much longer near zero intervals for non-anomalous points, as seen in Figure~\ref{fig:microcluster}. 

\subsubsection{Other (nearly) flat parts} \label{sub:otherflatparts} Similar to the (near) zero intervals, the DQFs can also display other (nearly) flat portions, where the increase of the DQFs is almost zero over a range of $\delta$ values not close to zero. On a heuristic level, this again indicates the presence of `empty regions' that are visible when looking in the `right' direction. Note, however, that here the flat parts are not caused by the behavior locally around the base point. Instead, assuming that the DQFs $q_{x,y}(\delta)$ for $\delta$-values in the region of interest are determined by the $A$-sets, say, i.e. $q_{x,y}(\delta) = F(A_{x,y})$, a flat spot at larger values of $\delta$ corresponds to a small change of mass of the $A$-sets when changing the cone tip, and for a larger value of $\delta$, the $A$-sets also become large. Recalling that, for fixed $(x,y)$, the $A$-sets are nested, we see that a flat spot of the DQF corresponds to an empty region along the surface of the corresponding $A$-sets. One example of a second flat spot in the DQFs can be seen in the green DQFs in Figure~\ref{fig:mfeat4v5} corresponding to the `5's.  

\begin{figure}
    \centering
\resizebox{4in}{!}{

\tikzset{every picture/.style={line width=0.75pt}} 

\begin{tikzpicture}[x=0.75pt,y=0.75pt,yscale=-1,xscale=1]

\draw   (2,1) -- (520,1) -- (520,294) -- (2,294) -- cycle ;
\draw  [fill={rgb, 255:red, 155; green, 155; blue, 155 }  ,fill opacity=0.36 ][line width=1.5]  (167.79,210.69) -- (300,120) -- (327.88,219.38) -- cycle ;
\draw [line width=1.5]    (300,120) -- (424,40) ;
\draw [line width=1.5]    (327.88,219.38) -- (468,227) ;
\draw [line width=1.5]  [dash pattern={on 1.69pt off 2.76pt}]  (417,45) -- (462,16) ;
\draw [line width=1.5]  [dash pattern={on 1.69pt off 2.76pt}]  (454,226) -- (515,229) ;
\draw  [dash pattern={on 0.84pt off 2.51pt}]  (10,253) -- (492,123) ;
\draw  [dash pattern={on 4.5pt off 4.5pt}]  (37,246) -- (375,14) ;
\draw  [dash pattern={on 4.5pt off 4.5pt}]  (37,246) -- (451,274) ;
\draw [line width=1.5]  [dash pattern={on 1.69pt off 2.76pt}]  (286,74) -- (300,120) ;
\draw [line width=1.5]  [dash pattern={on 1.69pt off 2.76pt}]  (327.88,219.38) -- (340,266) ;

\draw (229,170) node [anchor=north west][inner sep=0.75pt]   [align=left] {x};
\draw (222,172) node [anchor=north west][inner sep=0.75pt]   [align=left] {x};
\draw (381,173) node [anchor=north west][inner sep=0.75pt]   [align=left] {x};
\draw (285,130) node [anchor=north west][inner sep=0.75pt]   [align=left] {x};
\draw (186,77) node [anchor=north west][inner sep=0.75pt]   [align=left] {x};
\draw (363,76) node [anchor=north west][inner sep=0.75pt]   [align=left] {x};
\draw (341,185) node [anchor=north west][inner sep=0.75pt]   [align=left] {x};
\draw (427,197) node [anchor=north west][inner sep=0.75pt]   [align=left] {x};
\draw (381,79) node [anchor=north west][inner sep=0.75pt]   [align=left] {x};
\draw (451,160) node [anchor=north west][inner sep=0.75pt]   [align=left] {x};
\draw (410,125) node [anchor=north west][inner sep=0.75pt]   [align=left] {$\displaystyle \ell _{i}{}_{j}$};
\draw (16,235.4) node [anchor=north west][inner sep=0.75pt]   [align=left] {$\displaystyle x_{i}$};
\draw (16,247) node [anchor=north west][inner sep=0.75pt]    {$\bullet $};
\draw (169.79,214.09) node [anchor=north west][inner sep=0.75pt]    {$s$};
\draw (249,190.4) node [anchor=north west][inner sep=0.75pt]    {$A_{i}{}_{j}( s)$};
\draw (323,120.4) node [anchor=north west][inner sep=0.75pt]    {$B_{i}{}_{j}( s)$};
\draw (158,130) node [anchor=north west][inner sep=0.75pt]   [align=left] {x};
\draw (184,97) node [anchor=north west][inner sep=0.75pt]   [align=left] {x};
\draw (201,84) node [anchor=north west][inner sep=0.75pt]   [align=left] {x};
\draw (84,256) node [anchor=north west][inner sep=0.75pt]   [align=left] {x};
\draw (210,262) node [anchor=north west][inner sep=0.75pt]   [align=left] {x};
\draw (234,261) node [anchor=north west][inner sep=0.75pt]   [align=left] {x};
\draw (207,112) node [anchor=north west][inner sep=0.75pt]   [align=left] {x};
\draw (28,252.4) node [anchor=north west][inner sep=0.75pt]    {$s'$};
\draw (325,13) node [anchor=north west][inner sep=0.75pt]   [align=left] {x};
\draw (189,202) node [anchor=north west][inner sep=0.75pt]   [align=left] {x};
\draw (383,41) node [anchor=north west][inner sep=0.75pt]   [align=left] {x};
\draw (399.94,226.19) node [anchor=north west][inner sep=0.75pt]   [align=left] {x};
\draw (413,223) node [anchor=north west][inner sep=0.75pt]   [align=left] {x};
\draw (312,148.4) node [anchor=north west][inner sep=0.75pt]    {$m_{i}{}_{j}$};
\draw (310,166) node [anchor=north west][inner sep=0.75pt]    {$\bullet $};

\end{tikzpicture}
}
\caption{Illustrates the geometry underlying flat parts of $\wh q_{ij}(\delta)$; such flat parts are generated by flat parts of the function $d_{ij}(s)$, meaning stretches of cone tips $s$ without any change in the depth $\min(F_n(A_{ij}(s),B_{ij}(s));$ in the figure, the depth is determined by the number of points in the shaded area $A_{ij}(s)$, and because of the indicated void of data points, this number does not change when the cone tip moves from $s$ to $s'$, resulting in a flat part of $d_{ij}(s)$ and thus in a flat part of $\wh q_{ij}(\delta)$ of length being the $G$-measure of the interval $[s',s]$.}
\end{figure}


Whether or not such plateaus indicate `anomalies' in general is not entirely clear. Of course, any features of the DQFs that are displayed by only a few of them (whether flat regions or not) raise suspicions of anomalies, or, at least, they might indicate some interesting geometric features in the data that might be worth exploring further. Again, this is an advantage of a visualization tool.

\subsection{Choice of $\mathbf G_{\mathbf {x,y} } $ -  The Adaptive DQF} \label{adaptive}
In order to pronounce the discriminatory effect of the (near) zero intervals of the DQFs that has been discussed above (cf. type 1 zero intervals), we now introduce the adaptive DQF. The adaptation of the DQF approach is achieved via the function $G_{x,y}$, which is one of the ``tuning parameters." Again the discussion is presented in the population setting.

The formal results presented above, i.e., Lemma~\ref{basic}, Corollary~\ref{DQF-atzero} and Lemma~\ref{DQF-atzero2} provide important insights into the role of the distribution $G_{x,y}$ in the context of anomaly detection. Consider  depth quantile functions $q_{x,y}(\delta)$ with base points $m_{x,y}$. 
Our results imply that, for points $x$ such that $m_{x,y}$ is close to an antimode, or even fall outside the support, $q_{x,y}$ will tend to exhibit a (near) zero interval for values of $\delta$ close to zero.  On the other hand, if $m_{x,y}$ lies in the bulk of the data, then the slope of $q_{x,y}$ for small $\delta$ will tend to be large. This interpretation is mainly motivated by the magnitude of $f(m_{x,y})$. However, our results tell us that one actually should consider the magnitude of the ratio $\frac{f}{g}(m_{x,y})$ instead. The goal of adaptation now is to attempt to choose $G_{x,y}$ depending on $m_{x,y}$ in order to pronounce the discriminatory effect of the DQF described above, where we have type 1 anomalies in mind, i.e. $g_{x,y}$ should be (relatively) large for anomalous points $x$ and (relatively) small for non-anomalous points near $m_{x,y}$. To put it differently, the concentration of $G_{x,y}$ about $m_{x,y}$ should be pronounced when $x$ is an anomalous point. 

{\em Adaptive uniform distributions.} 
Here we propose to choose $G_{x,y}$ as the uniform distribution on $\ell_{x,y}$ over the range $[a_{x,y},b_{x,y}]$ of the push-forward distribution under the projection of $F$ onto $\ell_{x,y}.$ In practice, we of course use the empirical analog (projecting the empirical distribution onto $\ell_{x,y}$). This choice can be motivated as follows:

Consider again a contamination model on $\R^d$ of the form $(1 - \epsilon) P + \epsilon Q,$ where $\epsilon > 0$ is ``small", with $P$ a uniform distribution on the support $S_P$, and $S_Q \cap S_P = \emptyset.$  Suppose we use $\frac{x+y}{2}$ as the anchor point. For $X \sim P$ and $y \in S_Q$, the gap between $S_P$ and $y$ is measured by $\delta_{X,y}$, the length of the zero-interval of $q_{X,y}$ from Lemma~\ref{DQF-atzero2}. Notice, however, that the same physical gap (Hausdorff distance between $S_Q$ and $S_P$) can lead to different distributions of $\delta_{X,y}$, depending on where $S_Q$ is located under an adaptive choice of $G_{x,y}$. For instance, suppose that $S_P$ is an elongated ellipsoid, and $S_Q$ is a ball with Hausdorff distance to $S_P$ being a given positive value. Then, using a similar rationale as described above, we can see from Lemma~\ref{DQF-atzero2} that with this adaptive choice of $G_{x,y}$, the values of $\delta_{x,y}$ will tend to be smaller if $S_Q$ is located along the main axis of $S_P$ than if it is located along one of the minor axis. The reason is that the range of the corresponding push-forward distributions tends to be larger in the former case, and $\delta_{X,y}$ is the physical length of the gap divided by the range. In this latter case, this will lead to small values of the averaged DQF, $\overline{q}_{y}(\delta), y \in S_Q$ for a larger range of $\delta$. The rational for this adaptive choice is similar to that of  Mahalonobis distance, as a point at a fixed distance from $S_P$ in the direction of a minor axis is arguably more of an anomaly than one along the main axis. The same rational holds if we have a non-uniform distribution on $S_P$. 

{\em Adaptive normal distributions:} An adaptive choice for $G_{x,y}$ showing a strong performance in our simulation studies is a normal distribution on $\ell_{x,y}$ with mean $m_{x,y}$ chosen as the midpoint, and variance equal to a robust estimate of the variance of the push-forward distribution of the projection of $F$ onto $\ell_{x,y}$, such as a winsorized variance. The heuristic idea is that while winsorizing always reduces variance, the reduction in variance tends to be larger when an anomalous $x$ still tends to be anomalous for the push-forward distribution on the lines $\ell_{x,y}.$ The above discussion then shows that this leads to a longer (near) zero interval, which enhances the anomaly detection performance based on the DQF plots. As an extreme case, consider data living on a plane with an anomalous observation $x$ living off of it, and suppose that $y$ is such that $\ell_{x,y}$ is orthogonal to the plane.  By using a robust measure of spread, $G_{x,y}$ will be degenerate (the push-forward distribution is simply two  points, one with mass $\frac{1}{n}$) and thus $q_x(\delta)=0$ for all $\delta$.  In less extreme situations, the parallels to Mahalonobis distance discussed above also hold for this choice of base distribution.

Of course non-normal choices are also possible, including distributions with heavier tails such as $t$-distributions, etc.  

{\em Non-adaptive $G_{x,y}$:} For $G_{x,y}=G$, altering the base distribution from uniform simply amounts to a non-linear rescaling of the horizontal axis in the DQF plot, and accordingly only uniform base distributions were considered in Chandler and Polonik (2021).  Considering the above motivation for adaptation, one can imagine the non-anomalous support $C$ being a sufficiently curved manifold such that midpoints of non-anomalous pairs of points also tend to live outside the support $C$ and anomalous observations do not have non-anomalous projections.  In such cases, winsorizing will not have the above stated benefit. Even without clear geometric adaptivity, the functions are still likely to reflect the difference in geometry between anomalous and non-anomalous points, though visually the discriminatory information may exist away from $\delta=0$ (see Figures \ref{fig:microcluster} and \ref{fig:2manifold}), further motivating consideration of the entire function. 
%
 %

In the following we refer to a DQF approach with an adaptive choice of the distribution $G_{x,y}$ as the ``adaptive DQF'' (aDQF). We will make the actual choice of $G_{x,y}$ clear when presenting our numerical studies.

\subsection{Choice of the opening angle $\boldsymbol \alpha$}

Similar to the distribution $G_{x,y}$, the choice of the opening angle of the cones influences the shape of the (a)DQFs, and a good choice will depend on the underlying geometry and also on the dimension. As for the dimension, for a symmetric cone in $\R^d$ with a given height $h$ and radius of the base $r,$ the percentage of the volume of the sphere $S^d$ centered at the tip of a cone that is covered by the cone does not tend to zero as $d \to \infty$, if $\frac{\pi}{4} - \alpha = O(\frac{1}{\sqrt{d}})$.This knowledge is of limited use, however, as, for instance, the data often lies on a lower dimensional manifold. For practical application everything will depend $c$ when choosing $\alpha = \frac{\pi}{4} - c\frac{1}{\sqrt{d}}$, and again a good choice of $c$ will depend on the underlying geometry. 

Now, while a certain robustness to the choice of $\alpha$ has been observed in our numerical studies, a more refined study of the aDQF plots indicates subtle, but (at least visually) important differences for different angles. Our recommendation here is to consider a few choices of $\alpha$ at once and to compare the resulting plots. Computationally, this does not pose a huge burden, because in order to determine whether a point $w$ falls inside a cone with tip $s$ and axis of symmetry $\ell_{x,y}$, all we need to determine is the angle between $\ell_{x,y}$ and $\overline{sw}$. Then one simply compares this angle to the opening angle. Accordingly, the standard output of the R package presents the aDQF at three different values of $\alpha$.

\section{A connection to the shorth in the case ${\mathbf d}\; {\boldsymbol =}\; {\mathbf 1} $}
For anomalies of the type that are  far from the bulk of the data, techniques such as the box plot are both well-known and effective.  Instead, we consider the density based definition of an anomaly for $d=1$. We look at the case where anomalous observations exist in particular types of antimodes, which we call ``holes", characterized by a low density region surrounded closely by much higher density regions.  Such observations are occasionally referred to as ``inliers" in the literature (for example, Talagala et al., 2021).  As we relate distance based outliers to antimodes via the use of midpoints of pairs of points as our anchor points in the multidimensional case, this section also provides insight into the strength of the DQF approach for $d>1$.

\begin{figure}[h]
    \centering
    \includegraphics[height=2.5in,width=5in]{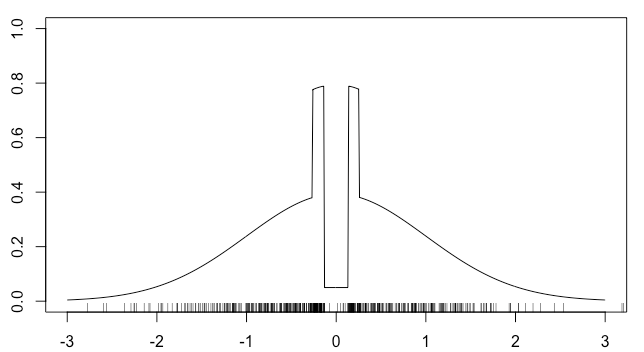}
    \caption{The density $f$ defined in (\ref{fn:normhole}) with a rug plot of a random sample of size $n=500$.}
    \label{fig:holeynorm}
\end{figure}

We consider the distribution used in Einmahl et al. (2010a), visualized in Figure \ref{fig:holeynorm} and defined as: 
\begin{equation}
f_X(x) = \label{fn:normhole}
  \begin{cases}
0.05 & \text{for}\quad |x| \leq 0.13 \\
 0.987[\varphi(x) + \varphi(|x| -0.13)] & \text{for }\quad 0.13 < |x| < 0.26 \\
0.987\varphi(x) & \text{for}\quad |x| \geq 0.26,
\end{cases}
\end{equation}
where $\varphi$ is the density of the standard normal distribution.

Einmahl et al. (2010a) motivated the {\it shorth plot} in part by demonstrating the shortcomings of standard techniques like the histogram or kernel density estimation in distinguishing this density (and its antimode) from a standard normal density based on a random sample.  Just as the DQF $q_x(\delta)$ is related to the probability measure of an interval of size $\delta$ for which $x$ is the ``median'', the shorth, $S_{\lambda}(x)$, also relates the length of a particular interval with its probability content, specifically 
$$S_{\lambda}(x) = \inf\{|I|:P(I)\geq \lambda, I\in \mathcal{I}_x\},$$ where $\mathcal{I}_x$ is the class of intervals containing the point $x$.  A similar approach was considered by Lientz (1974), where the set ${\cal I}_x$ was the set of intervals with {\em midpoint} $x$. Empirical versions are found by again replacing the probability measure with its empirical counterpart.

\begin{figure}[h]
    \centering
    \includegraphics[scale=0.25]{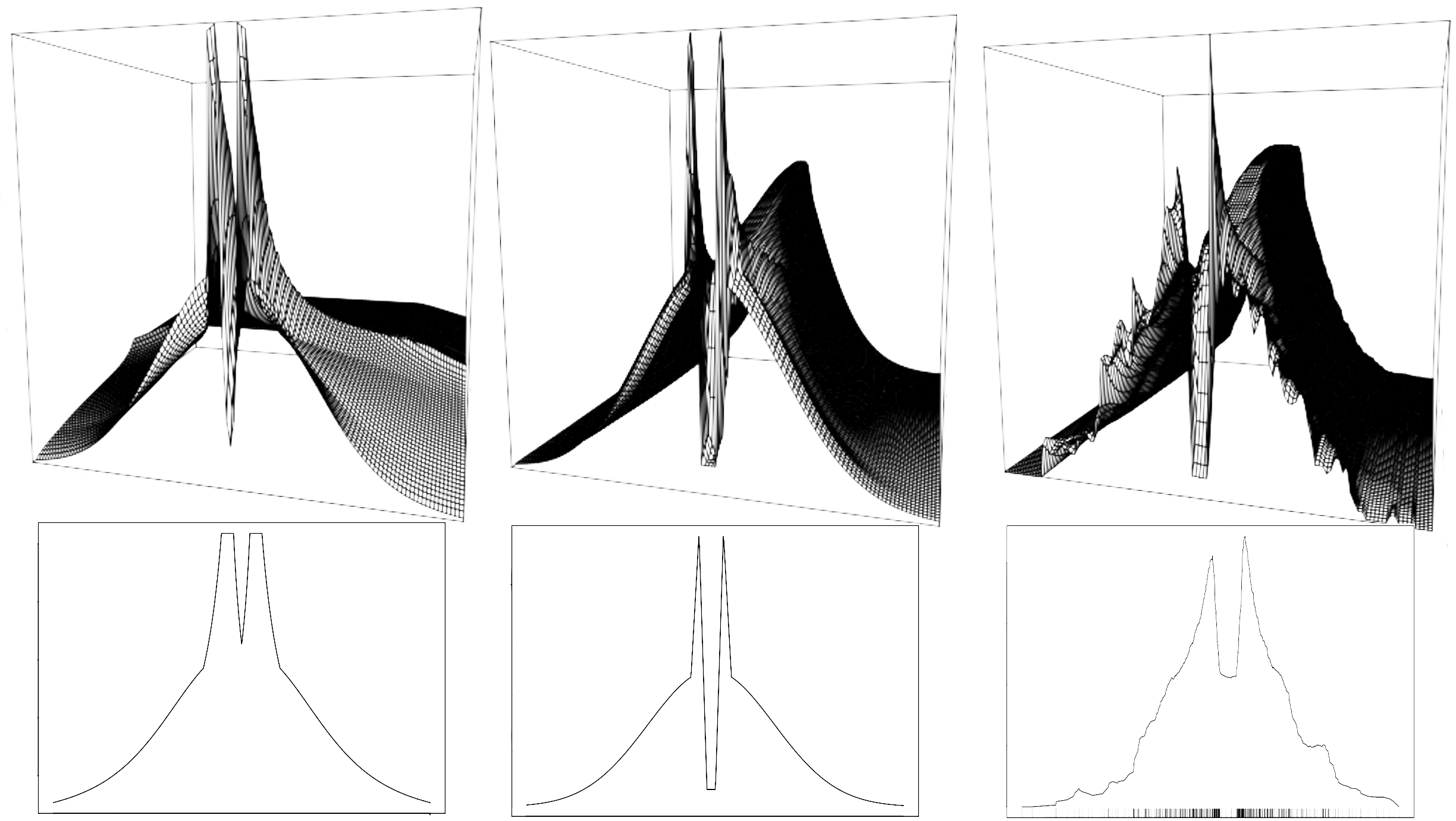}
    \caption{(left) reciprocal of the shorth over all $\lambda$; (middle) DQF over all $\delta;$ (right) estimated DQF based on a sample of size $n=500$, smoothed over $\delta$. All have been scaled at each $\lambda(\delta)$ to aid in visualization. Bottom shows slices at individual values of the scale parameter.} 
    \label{fig:holynorm_emp}
\end{figure}

The shorth-plot was introduced as a function in $x$ for a given parameter $\lambda$ in Sawitzki (1994). The shorth-process considered in Einmahl et al. (2010b) is studied as a process in both $x$ and $\lambda$. The DQF is defined as a function in the parameter $\delta$ for each fixed $x,$ but, for $d=1,$ it can also be plotted as a function in $x$. Figure \ref{fig:holynorm_emp} provides plots of these functions in both $x$ and the corresponding parameter. To make them comparable, we plot the reciprocal of $S_\lambda(x)$ (left figure), in addition to normalizing each plot such that, for a fixed value of the scale parameter, the integral over $x$ is 1. This means that for small parameter values, both functions (in $x$) might be considered as some type of density estimator. However, neither of these approaches is to be understood as such. Indeed, particularly in high dimensions when the curse of dimensionality kicks in, we are aiming at extracting more structural information in the sense of feature extraction.

One notable feature is that, in the low density ``hole", the shorth is ``v"-shaped, while the DQF (both population and smoothed empirical versions) is ``u''-shaped, matching the underlying density. To see why this is, consider the point $x=0.10$, which exists in the ``hole", though near its right hand boundary of 0.13.  The interval of length 0.15 for the DQF will be approximately (-0.015, 0.135), almost entirely in the ``hole," and something similar will be true for any $x$ value in the ``hole."   Meanwhile, the interval of the same length that will be considered by the shorth will be (0.1, .25) (corresponding to $\lambda$=0.095).  Thus, as $x$ nears the boundary of the ``hole," the intervals instead increasingly resemble the intervals of non-anomalous points, masking the hole and resulting in the observed ``v" shape. 

A second feature of Figure \ref{fig:holynorm_emp} is that, for large $\lambda$, the information quality tends to degrade as $\lambda\to 1$, since $S_1(x) = |\supp(F)|$ for all $x$.  This is in contrast to the DQF, which, by Lemma~\ref{basic} yields information about the centrality of $x$ for large $\delta$.  While 
visualizations of the type contained in Figure \ref{fig:holynorm_emp} will not be possible beyond the univariate setup, the realization that valuable information is contained at both extremes of $\delta$ motivates plotting the DQF for each data point as a function of $\delta$, which will be possible for data of any dimension, as seen below.


\section{Anomaly detection for $\mathbf{d}\boldsymbol{\geq} \mathbf{2}$ and object data}
In this section numerical studies of the (a)DQF approach to anomaly detection are presented and discussed. 
We fix the cone angle $\alpha = \pi/4$ unless mentioned otherwise. We also drop the axis from the plots of the DQFs and their derivatives, because the scale is not of importance for our discussion. Recall that as described in subsection~\ref{adaptive} (see Uniform distribution (iii)), the non-adaptive DQF uses the same (uniform) distribution $G_{ij}$ for all $i,j$. 
\subsection{Anomaly detection for multivariate Euclidean data}
In low dimensional situations, using the observations themselves as anchor points may make sense.  For instance, this would be beneficial in trying to detect the hole in a bivariate ``rotated'' version of (\ref{fn:normhole}). However, as the dimension of the data increases, issues related to the curse of dimensionality arise.  A particularly salient one for the current method is that the fraction of points living on the boundary of the convex hull increases in dimension (R\'{e}nyi and Sulanke, 1963).  Points living on this boundary will tend to have very small half space depth. As half space depth underlies the construction of the depth quantile functions, in high dimensions the DQF for a large quantity of points will have nearly identical behavior, which empirically has been seen to be functions that only take on the values 0 and $1/n$. Thus, the ability to recognize anomalous observations is hampered.  This provides rational for using $m_{ij}=\frac{x_i+x_j}{2}$ as our anchor points. Beyond having the aforementioned benefit of relating to antimodes for anomalous observations, these points are very likely to live in the interior of the convex hull and thus return informative (a)DQFs.  Each observation is then associated with the averaged (a)DQF over midpoints formed between itself and other observations. Considering midpoints has the additional benefit of reducing the computational complexity of the algorithm, from $n(n-1)$ comparisons when the observations themselves are the ``anchor points" to ${n\choose 2}$ when using midpoints.  In the applications that follow, the averaged (a)DQFs are based on a random sample of 50 pairs rather than all pairs of points, further reducing the computational burden of the method.  

A common feature of high dimensional data is the so-called manifold hypothesis (Fefferman et al. 2016) that posits that the data often lives (at least approximately) on a manifold of much lower dimension than the $d$-dimensional ambient space.  This fact is often exploited by dimension reduction algorithms, either linear (for instance, principal component analysis) or non-linear (for instance, Isomap by Tenenbaum et al. 2000).  Certainly an observation far in the tails of the point cloud might be considered anomalous, and this is the higher dimensional analog of what the box plot is designed to detect. Considering a definition of an outlier as an observation that is generated according to a different mechanism from the rest of the data set, such a point may lie off of the manifold, despite otherwise living ``near'' the point cloud (with respect to the ambient space).

A particular benefit of the DQF approach is adaptivity to sparsity.  Suppose that the non-anomalous data lives in a $d'$ dimensional affine subspace of the ambient space with dimension $d$, $d'<d$.  For any two non-anomalous points, the line $\ell_{ij}$ will also live in this subspace, and the intersection of our $d$-dimensional cones with this subspace will be cones of dimension $d'$.  In other words, the DQF approach will behave as if we had done appropriate dimension reduction beforehand. 

In the following empirical studies, we use midpoints as anchor points, i.e. $m_{ij} = \frac{x_i + x_j}{2}.$  The aDQF uses a normal base distribution, centered at $m_{ij}$ with variance proportional to the $(\frac{n-6}{n})100\%$-winsorized variance of the projected (onto $\ell_{ij}$) data. In the Euclidean case, the data is first $z$-scaled.  

While the DQF approach results in a functional representation of each observation, a particularly informative region of these functions is for $\delta$ small, particularly the range of $\delta$ for which $\bar{q}(\delta)$ is 0. Considering the zero interval of the (a)DQFs means to look for 
``large distance gaps,'' measured by $\delta_{x,y}$ (see Lemma~\ref{DQF-atzero2}). Of course, Lemma~\ref{DQF-atzero} also makes clear that $\delta$ small is related to the underlying density, resulting in somewhat of a unification of the two standard types of outliers, those far from the bulk of the data and those lying in low-density regions, even in the case of micro-clusters.

To explore the sense in which the DQF approach to measuring distance gaps can facilitate anomaly detection, we contrast the DQF approach with a selection of anomaly detection algorithms found in the literature.  We consider the one-class support vector machine  (OCSVM) introduced by Sch\"{o}lkopf et al. (2001), which attempts to find the support of the underlying distribution.  With the introduction of slack variables, it is possible to handle anomalies in the training data. Negative decision values correspond to points outside the learned support.  As we are using a notion of depth in the current method, we consider a classical depth measure for anomaly detection.  While many depth measures, like Tukey depth, vanish at the convex hull, Mozharovskyi and Valla (2025) show that projection depth can be used for anomaly detection.  Fair Cut Forests (Cortes, 2021) is a powerful variant of Isolation Forests (Liu et al. 2008).  It is based on the idea that outliers can be isolated by relatively few binary tree splits. Finally, {\it stray} (Talagala et al., 2021) is an extension of the {\it HDoutliers} algorithm (Wilkinson, 2017),  a well-studied and popular anomaly detection method based on the distribution of the $k$-nearest neighbors distances, specifically for anomalies defined by large distance gaps. Using neural nets for anomaly detection dates back at least to the work of Bishop (1994). More recently, autoencoders are used for this task.  The idea is that for anomalous observations, the autoencoder will have difficulty reconstructing these observations.  Accordingly, outliers are identified as having large residuals, though it has been demonstrated that this need not be true, see  e.g. Bouman and Heskes (2025). Though heuristics exists, a data set consisting entirely of non-anomalous observations is typically required to train the autoencoder, whereas the aforementioned techniques do not have this requirement.  Therefore, autoencoders are not considered in our numerical simulations.    

As we are in an unsupervised learning setting, model tuning is difficult.  Accordingly, we run all comparison algorithms using their default parameters in their R implementation: OCSVM  from {\it e1071} (Meyer et al. 2022), projection depth from {\it fdaoutlier} (Ojo et al., 2023), FCF from {\it isotree} (Cortes, 2025) and stray from {\it stray} (Talagala, 2020). 

We note a handful of issues that our comparison methods may suffer from. For classical depth measures, star-convexity of the depth contours is a required property, suggesting that the method may struggle with non-convex supports of the non-anomalous distribution. In the case of stray, with an underlying manifold structure, an outlier lying off the manifold may have a small distance gap to other observations with respect to Euclidean distance in the ambient space, but a large distance gap with respect to a Mahalanobis-type distance with respect to the distribution on the manifold.  We argue that the DQF approach, and particularly the aDQF, is sensitive to this type of anomaly and implicitly recognizes the large distance gap relative to a notion of distance based on the underlying geometry without the need for manifold learning.
\subsubsection{Simulation Study}
\begin{figure}[h]
    \centering
    \includegraphics[height=2in, width=4.5in]{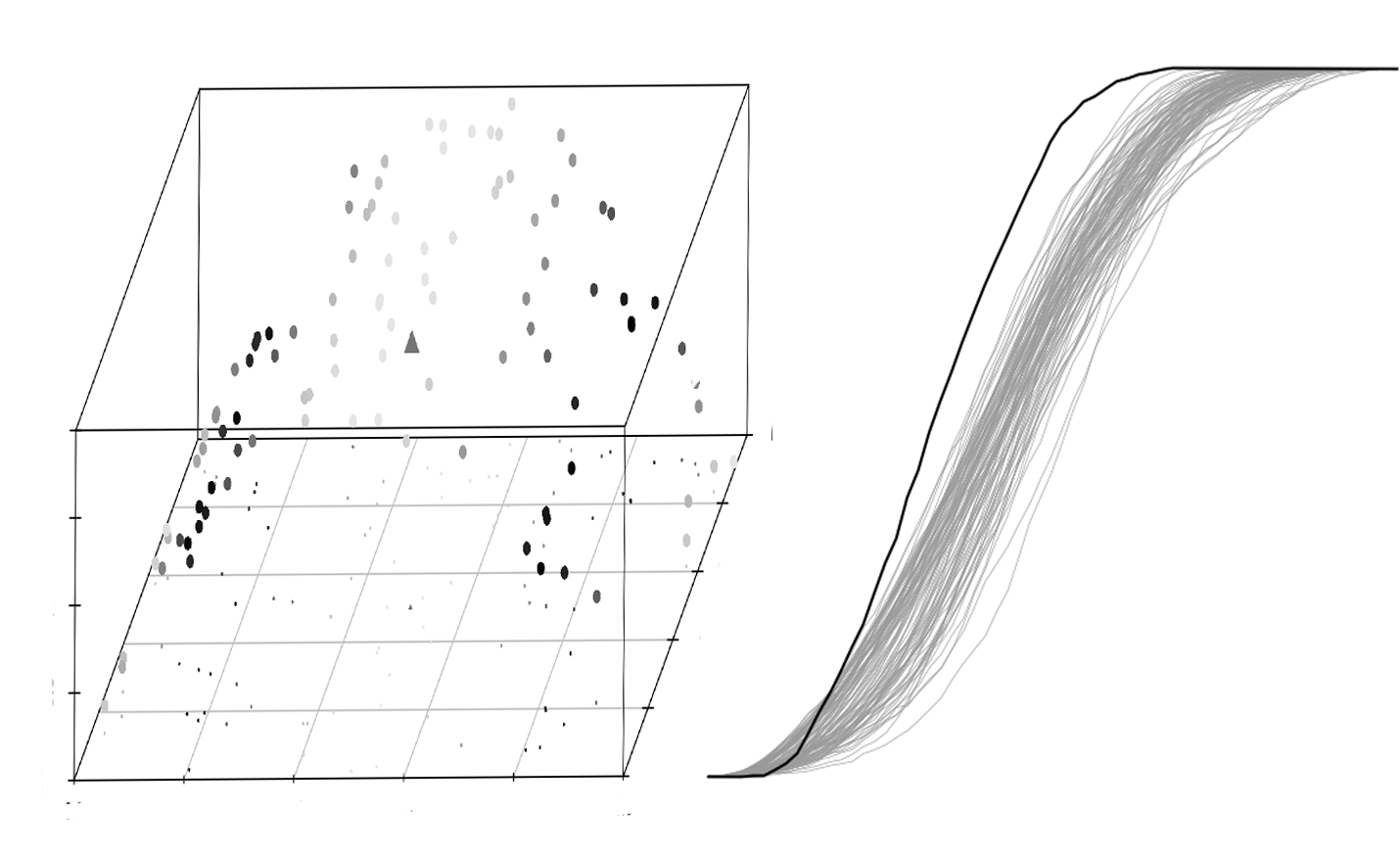}
    \caption{$x_1,\ldots,x_{100}$ living on a 2-manifold embedded in $d=3$ with $x_{101}$ living off the manifold. (left) Data, color coded by vertical height (light to dark to light), outlier as triangle 0.5 below the apex of manifold, with projection plot;  (right) Normalized aDQFs $\tilde{q}_i(\delta)$ with outlier in black.}
    \label{fig:2manifold}
\end{figure}

We consider a series of illuminating simulation studies to contrast the various anomaly detection algorithms.  For the first, we generate $n=100$ observations living on a 2-manifold defined by $y_3=2\cos((y_1-.5)\pi)$ with $(Y_1,Y_2)\sim U(0,1)^2$. An additional observation lives at the point (0.5,0.5,1.5), so that the manifold is curving around this observation.  Such a value could come about by someone unscrupulously creating a synthetic observation by taking a convex combination of two existing values.  Results are provided in table \ref{tab:curved_mani}. The OCSVM always gave the largest decision value to the anomaly, suggesting that the estimated support of the distribution was built around the outlier.  Similarly, projection depth is unable to deal with the curvature of the manifold about the outlier and regularly finds the observation to be very deep in the point cloud. 
Figure \ref{fig:2manifold} presents a simulated data set, 
as well as the corresponding normalized aDQFs $\tilde{q}_i(\delta)$'s. The function corresponding to the outlying observation differentiates itself from the bulk of the data, including but not limited to a long zero interval.  
\begin{table}
\centering
\begin{tabular}{|c|c|c|}
\hline
Method & Proportion of Correct Rank & Average Rank\\ \hline \hline
DQF & {\bf 0.61} & 2.93\\ \hline
aDQF & 0.58 & 3.09\\ \hline
Stray & 0.22 & {\bf 2.60}\\ \hline
PDepth & 0.00 & 98.41\\ \hline
FCF & 0.03 & 9.30\\ \hline
OCSVM &  0.00 &  101\\
\hline
\end{tabular} \label{tab:clean2mani}
\caption{Simulation results for a curved 2-manifold embedded in 3-space.}
\end{table}

Next,  we consider high-dimensional data with $n=80$ and a single anomaly.  We again generate $(Y_1, Y_2)\sim U(0,1)^2$ and then generate a random $50\times 2$ matrix $A$ with coefficients generated as Uniform(-1,1). We then consider $X=AY+\epsilon$, which roughly lives on a 2-dimensional affine subspace of a 50-dimensional space, where $\epsilon\sim {\rm MVN}(0,0.001^2I)$.  A single point is then shifted by 0.3 in a random orthogonal direction to the affine subspace, which constitutes the anomaly.  For either method based on the DQF, every midpoint involving the anomalous point lives off the manifold, and accordingly both methods are very good.  The Fair Cut Forest seemingly struggles with the dimension of the ambient space.  With pairwise distances at times larger than 5 in the ambient space, the small deviation from the manifold seems to affect the performance of the OCSVM.  

\begin{figure}[htbp]
  \centering
    \includegraphics[width=0.8\textwidth]{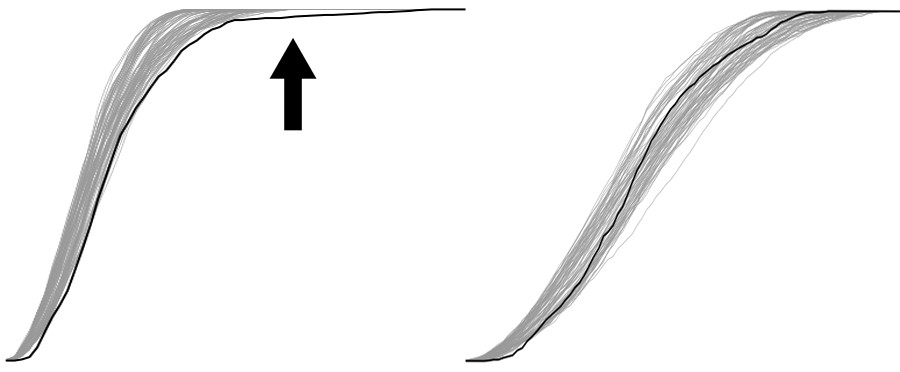}
  \caption{The right hand behavior (indicated by arrow) of the anomaly is much more informative under the aDQF (left) than the non-adaptive DQF (right) for non-anomalous data living approximately on an affine subspace.}
  \label{fig:planeAvsNot}
\end{figure}

As expected, the adaptive DQF outperforms the non-adaptive version, though visually the information provided is much more stark. As we can see in figure \ref{fig:planeAvsNot}, the adaptivity means that points for which $l_{xy}$ is nearly orthogonal to the plane, the base distribution will put almost all its mass in the gap, and the ``global'' depth will not be reached until very far in the tails.  As a result, the right-hand behavior of the aDQF is very different for the outlier than the non-anomalous points.  

\begin{table}
\centering
\begin{tabular}{|c|c|c|}
\hline
Method & Proportion of Correct Rank & Average Rank\\ \hline \hline
DQF &  0.92 &  1.29\\ \hline
aDQF & {\bf 0.95} & {\bf 1.24}\\ \hline
Stray & 0.33 & 3.69\\ \hline
PDepth & 0.76 & 3.01\\ \hline
FCF & 0.07 & 10.43\\ \hline
OCSVM &  0.06 & 18.31\\
\hline
\end{tabular} 
\caption{Simulation results for a linear 2-manifold and anomalous point shifted orthogonally from plane.}\label{tab:curved_mani}
\end{table}

\begin{figure}
    \centering
    \includegraphics[width=1\linewidth]{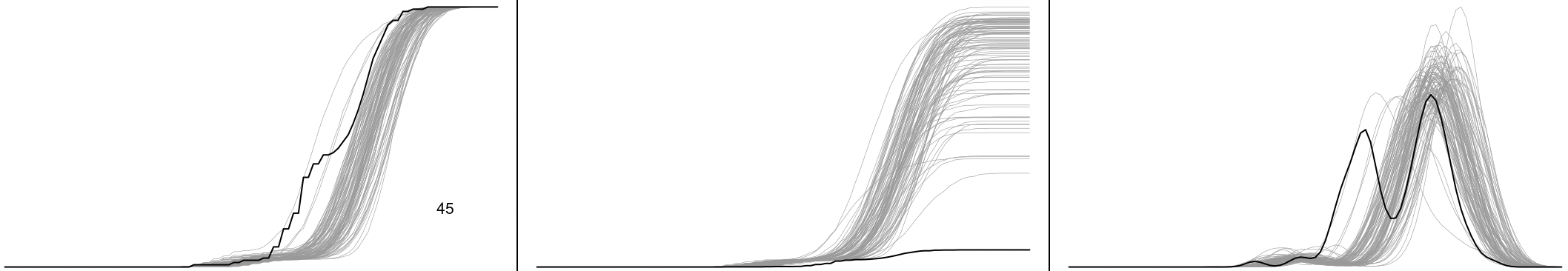}
    \caption{The three visualizations at angle $\pi/4$ for the aDQF for the simulation setup with points lying near a 6-d affine subspace embedded in 100 space. The generated anomaly is in bold. }
    \label{fig:noisysix}
\end{figure}

\begin{table}
\centering
\begin{tabular}{|c|c|c|}
\hline
Method & Proportion of Correct Rank & Average Rank\\ \hline \hline
DQF &  0.50 & 3.56\\ \hline
aDQF &  0.48 & 4.94\\ \hline
Stray & 0.00 & 99.5\\ \hline
PDepth & 0.26 & 7.89\\ \hline
FCF & 0.43 & 5.77\\ \hline
OCSVM &  {\bf 0.96} & {\bf 1.11}\\
\hline
\end{tabular} 
\caption{Simulation results for a noisy linear 6-manifold and anomalous point shifted orthogonally from plane.}\label{tab:noisy6}
\end{table}
Thirdly, we consider $n=100$ observations living near a 6-dimensional affine subspace embedded in 100-dimensional space via a random linear map $A$ generated similarly as above. After mapping to the ambient space, we add noise to each point of the form $\epsilon\sim N(0, .05^2 I).$  The final observation in the data set is then replaced with the first observation shifted by a random vector, orthogonal to the intrinsic space, with norm 10.  Due to distance concentration, these two observations are very likely to have the smallest pairwise distance.  As stray is based on $k$-nearest neighbors, it is incapable of identifying the anomaly.  With our evaluation criterion motivated by measuring the length of the zero interval, the results in table \ref{tab:noisy6} suggest substantial under-performance relative to the OCSVM.  However, for the outlier, all midpoints live away from the manifold with directions nearly orthogonal.  Accordingly, it is often the case that either $A(s)$ or $B(s)$, whichever contains the anomaly, is likely to be sparsely populated.  This makes visual identification of the outlier often readily apparent by considering the large scale values of the (a)DQF, as seen in the center panel of figure \ref{fig:noisysix}.

\begin{table}
\centering
\begin{tabular}{|c|c|c|}
\hline
Method & Proportion of Correct Rank & Average Rank\\ \hline \hline
DQF &  0.511 & 2.77\\ \hline
aDQF & 0.495 & 3.00\\ \hline
Stray & 0.019 & 9.46\\ \hline
PDepth & 0.27 & 4.87\\ \hline
FCF & 0.19 & 10.41\\ \hline
OCSVM &  {\bf 0.53} & {\bf 2.60}\\
\hline
\end{tabular} 
\caption{Simulation results for non-manifold data.}\label{tab:non_mani}
\end{table}
As the last simulation example, we consider a non-manifold example, with $n=50$ observations generated from a 30-dimensional standard normal distribution.  The first coordinate of one observation is then manually changed to 6, which constitutes the anomaly.   As there is no manifold structure to exploit, the aDQF performs nearly identically to the DQF, as seen in table \ref{tab:non_mani}.

\subsubsection{Real Data}
\begin{table}
\centering
\begin{tabular}{|c|c|c|c|c|c|c|c|c|}
\hline
& aDQF & DQF & ProjDepth & FCF & OCSVM & Stray\\ \hline\hline
ECG & {\bf 0.932} & 0.916 & 0.919 & 0.912 & 0.918 & 0.872\\\hline
mfeat: 0 vs All & {\bf 0.9923} & 0.9919 & 0.978 & 0.990 & 0.9916 & 0.991\\\hline
mfeat: 4 vs 7 & 0.949 & 0.986 & 0.993 & {\bf 0.996} & 0.995 & 0.762\\ \hline
Satellite & 0.719 & 0.773 & 0.759 & {\bf 0.867} & 0.759 & 0.833\\ \hline

\end{tabular}

\caption{Simulation results for real data sets, area under the ROC curve (AUC).}\label{tab:not_distance}
   
\end{table}
The first data set we consider is a collection of time series corresponding to electrocardiograms (ECG), considered by Chen et al. (2015).  The data set consists of $n$=5000 ECG readings with $d=140$ time points. The data is labeled, with each ECG corresponding to a normal or abnormal reading.  We created data sets by randomly sampling $n_1=50$ normal readings and $n_2=3$ abnormal readings, constituting anomalies.  The area under the receiver operating characteristic curve (ROC AUC) was then computed to evaluate the different methods.  The adaptive DQF yielded the best results in the 1000 simulations. 

\begin{figure}[h]
    \centering
    \includegraphics[scale=0.5]{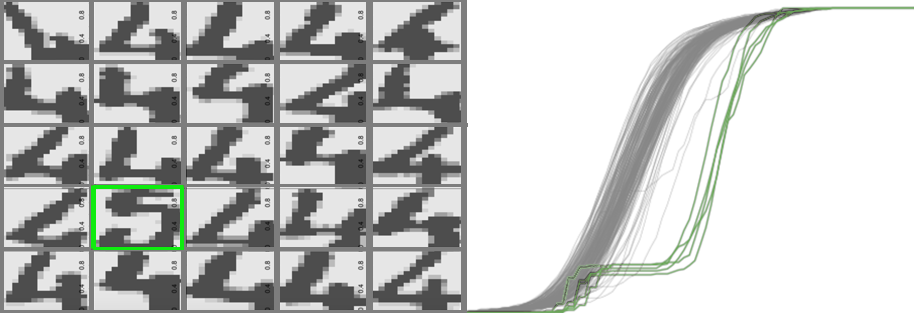}
    \caption{Normalized DQFs $\tilde{q}_i(\delta)$ from the {\it multiple features} data set.  200 ``4'' digits (red) and the first 5 ``5'' digits (green).}
    \label{fig:mfeat4v5}
\end{figure}

As a second real data example, we consider the {\it multiple features} data set available at the UCI Machine Learning Repository (\url{https://archive.ics.uci.edu/}).  The data set consists of $d$=649 features of handwritten digits, with 200 observations for each digit.  This data set was considered in Chandler and Polonik (2021), where a data set consisting of all ``6" and ``9" observations were contaminated with 20 ``0"s.  Figure \ref{fig:mfeat4v5} considers all 200 ``4'' digits and the first five ``5'' digits in the data set. The DQFs again use midpoints as anchor points. The 5 anomalous observations clearly separate themselves out from the rest of the data. FCF ranks the 5's as the five most anomalous points, whereas {\it stray} assigns ranks 2,3,4,10 and 11, 
OCSVM 2 through 6, and projection depth 1,2,3,4, and 8. For the anomalies,  flat parts away from $\delta=0$ are observed, as discussed in section \ref{sub:otherflatparts}. These are due to the five anomalies being the only observations in the $A$-sets for a large range of cone tips. This behavior also suggests that the manifold comprising the non-anomalous support curves around the five outliers, similar to what is illustrated in figure \ref{fig:zeroint}. The curvature of this manifold is seemingly the reason that the visual information in the non-adaptive DQF presented here is better than the aDQF. Visualizing some other observations displaying interesting functions reveals many are 4's that are upside down.  More extensive simulation studies are run comparing a data set consisting of $n_1=75$ observations from a non-anomalous class and $n_2=3$ anomalies.  The two situations considered are non-anomalies being the 0's and anomalies sampled from all other digits, as well as 4's vs 7's.  ROC AUC's for the 6 methods are contained in table \ref{tab:not_distance}.


Finally, we consider the Satellite data set, available in the R package {\it mlbench} (Blake and Merz, 1998).  It consists of  multi-spectral values of pixels.  Each of four spectral bands yields values in a $3\times 3$ grid, for a total of $d=36$ values per observation.  There are a total of 6 classes (soil types) and observations of types ``damp gray soil", ``cotton crop", and ``vegetation stubble"
are treated as outliers.  In our comparison study, we repeatedly generate data sets consisting of random samples of $n_1$=60 observations from the non-anomalous classes along with $n_2=4$ from the anomalous classes.  Fair class forests yield the best performance.  It is worth noting that if the evaluation criterion for the DQF approaches had instead been the value at the 80th percentile (at or near the ``global'' depth), then both methods return average AUCs of approximately 0.74.  The correlation between the AUCs of the  DQF at the 0.8 quantile and our standard early quantile is only 0.34. Though not entirely reasonable but mimicking consideration of the entire function, if we chose the higher AUC of the two quantiles for each simulation run, the DQF would yield an average AUC of 0.83.

\begin{figure}
    \centering
    \includegraphics[width=1\linewidth]{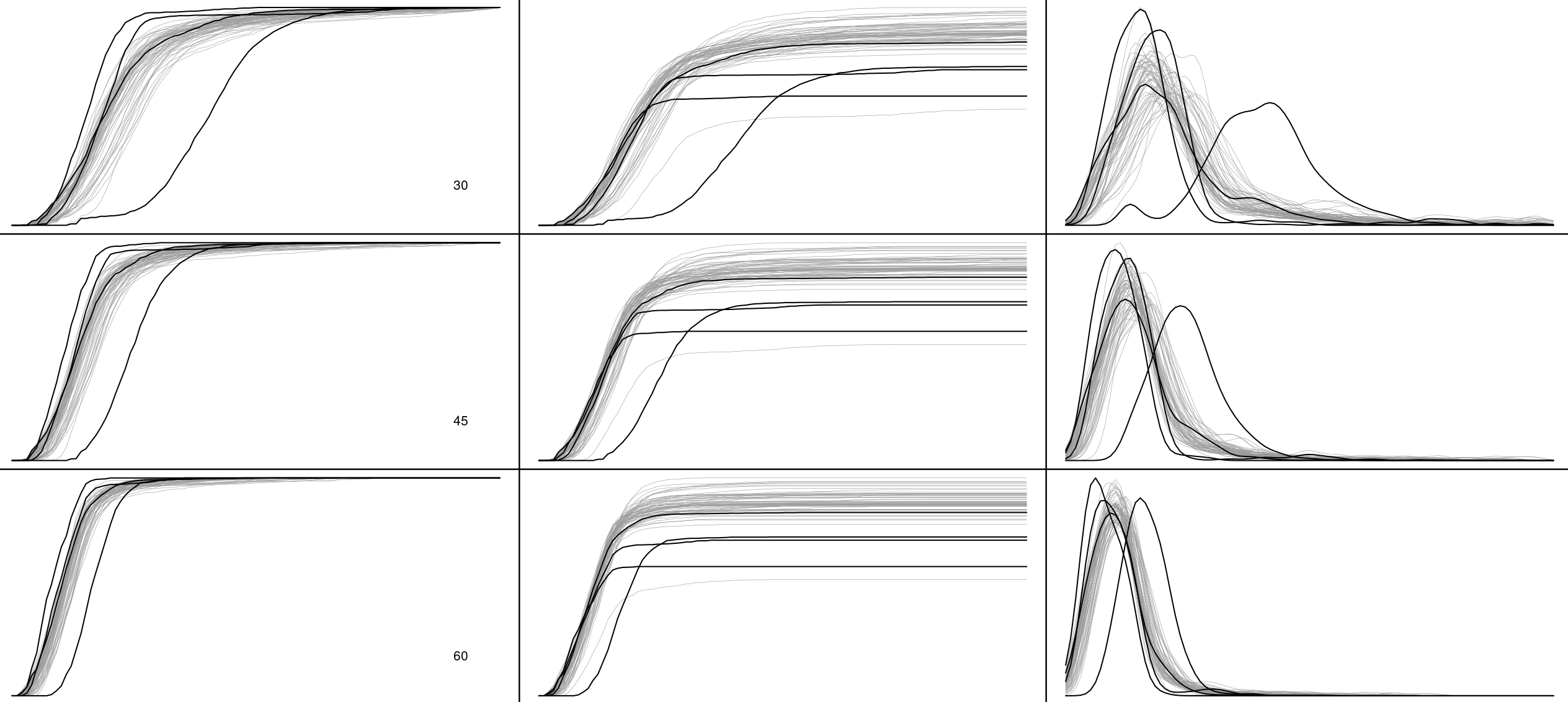}
    \caption{The output {\it dqf.explore} for a random sample from the Satellite data. The 4 observations from the anomalous classes are visualized.}
    \label{fig:satgraph}
\end{figure}

Figure \ref{fig:satgraph} visualizes a particular random sample from the simulation study. Three of the four anomalies are readily apparent, along with a fourth non-anomalous observation that appears to exhibit anomalous behavior (observation 1300 in the data set), labeled as belonging to class ``very damp gray soil'', one of 25 from this class in the data set. The final anomaly does not stand out, though its large-scale behavior ranks it $10^{th}$, while FCF and OCSVM rank it the $44^{th}$ and $54^{th}$ most unusual, respectively.  This is likely because it belongs to class ``damp gray soil'' ($n=1$), while ``gray soil'' ($n=15$) and ``very damp gray soil'' ($n=25$) are treated as non-anomalous classes here, meaning ``damp gray soil'' is likely an ``inlying'' anomaly. 
\subsection{Object Data - Kernelized DQF}
Given that the (a)DQF only requires computing distances between observations, any object data for which a kernel function is defined can be visualized by using the corresponding Reproducing Kernel Hilbert Space (RKHS) geometry via the (a)DQF.  Here we consider three such examples.  Any kernelizable anomaly detection algorithms, for instance OCSVM, could also be used here, though without a visualization aspect.

More recently, a related idea has been introduced, assuming only the existence of a metric.  Dubbed {\it metric statistics} (Dubey et al., 2024), each object is similarly related to a function of a single variable called the {\it distance profile}, which describes the distribution of distances from a given observation to a random observation in the space.  Thus, points with distance profiles shifted to the right would indicate outlyingness. From the viewpoint of the DQF, one can arrive at this idea by considering balls rather than cones, and not subdividing the balls into two sets.  Considering this idea with Euclidean data, it is clear that not exploiting the additional structure of the Hilbert space would cause suboptimal performance on many of the simulation studies above, though examples exist for object data living only in a metric space.  

\begin{figure}
    \centering
    \includegraphics[height=2in, width=5in] {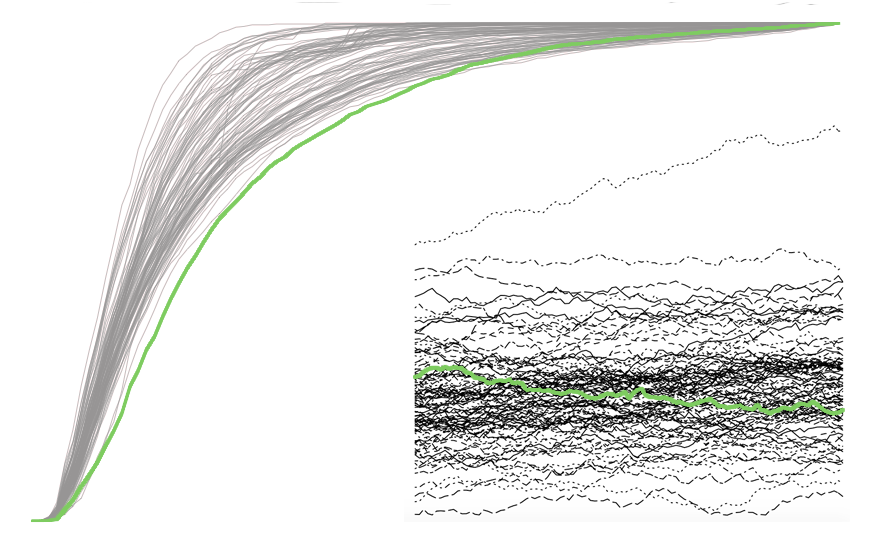}
    \caption{Functional data example, raw data (inset) and normalized aDQF $\tilde{q}_i(\delta)$. The anomaly (green) tends to decrease while the non-anomalous observations tend to increase.  The other observation identified by the aDQF corresponds to the highest function. }
    \label{fig:fnanom}
\end{figure}

In each of the following examples, the data itself can be visualized, unlike the previous Euclidean data settings. However, the (a)DQF, guided by Lemma~\ref{DQF-atzero2}, provides a visualization that structures the data according to its outlyingness, facilitating identification that may be difficult when observing the raw data.  We begin by considering Example 1 from Nagy et al. (2017).  We generate 100 functions of the form $X(t)=A + B \arctan(x) + G(t)$, with $A\sim N(0,4)$, $B\sim {\rm Exp}(1)$ and $G(t)$ a zero-centered Gaussian process with covariance function $k(s,t)=0.2 e^{\frac{-|s- t|}{ 0.3}}$, $0<s,t<1$. These constitute non-anomalous observations, while a  $101^{st}$ observation generated as  $Y(t)=1-2\arctan(t)+G(t)$ forms the anomaly, off the manifold bounded by $B=0$.  Figure \ref{fig:fnanom} shows both the raw data  and the aDQF based on the $L_2$-inner product.  

The second example is the shape classification task studied in Reininghaus et al. (2015), where we analyze 21 surface meshes of bodies in different poses, in particular the first 21 observations of the SHREC 2014 synthetic data set.  The first 20 correspond to one body, while the $21^{st}$ corresponds to a different body, which we regard as an outlier. 

\begin{figure}[h]
    \centering
    \includegraphics[scale=0.5]{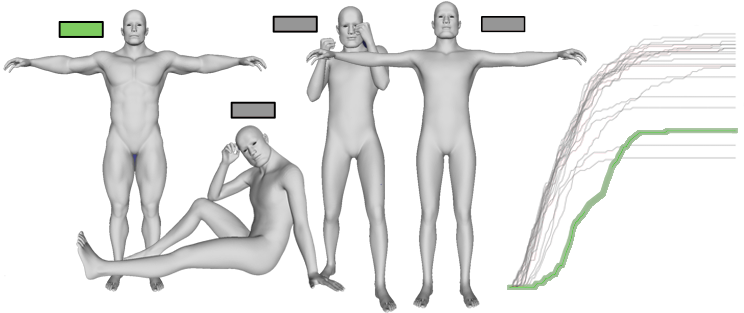}
    \caption{ aDQFs $\bar{q}_i(\delta)$ based on a kernel defined on persistence diagrams.  Red lines correspond to 20 poses of one body, green line corresponds to a single pose of a different body (outlier).  Tuning parameters of $t=500$ for the HKS and $\sigma=1$ for the PSSK were used.}
    \label{fig:TDA_dqf}
\end{figure}

The approach is based on the so-called heat kernel signature (HKS) defined on surface meshes introduced in Sun et al. (2009). In our application, we consider the HKS $h_t(x)$ as a function on the mesh, parameterized by a tuning parameter $t$. For small $t$, the HKS contains curvature information. The problem with using the HKS is that two HKSs corresponding to two meshes cannot be compared directly, because they are defined on different domains. Thus, one settles for comparing features of the HKSs. The idea put forward in Reininghaus et al. (2015), is to use topological features summarized in a persistence diagrams (e.g. see Chazal and Michel, 2021). As persistent diagrams are (2D) point sets (and thus not exhibiting a vector space structure), they themselves are then mapped to a feature space (RKHS) using the so-called persistent scale-space kernel (PSSK). For the interested reader, some background on the HKS, topological features and the PSSK are provided in the appendix. Ultimately, the RKHS geometry induced by the PSSK provides distances between feature space embedded persistence diagrams that serve as proxies for distances between the meshes. Using these pairwise distances in an RKHS, we are able to apply the aDQF methodology. The resulting aDQF plot shown in Figure~\ref{fig:TDA_dqf} provided stronger visual evidence than the non-adaptive DQF.    

Our third example in this subsection considers gerrymandering. We follow the techniques of Duchin et al. (2021) to compute a persistence diagram based on a graph representation of districting maps filtered by Republican vote percentage in the 2012 Pennsylvania Senate election. Again one can think of these persistence diagrams being constructed by first constructing a function over each potential districting map. This is accomplished by associating each districting map with a graph (vertices are the districts; edges are placed between vertices if the districts physically share a part of their boundary). Then, a function is constructed over this graph, by defining a function value at each vertex (district), which is given by the portion of Republican votes (based on counts of precincts at a given election). Then, as in our previous example, one constructs a persistence diagram for this function by again using its level set filtration. This is done for both the 2011 map invalidated by the Supreme Court for partisan gerrymandering as well as random maps generated according to the ReCom algorithm (Deford et al., 2021). The idea is that these randomly chosen districting maps constitute a representative sample of maps, and they are being used to answer the question of whether the gerrymandered map is an anomaly. 

\begin{figure}[h]
    \centering
    \includegraphics[scale=0.5]{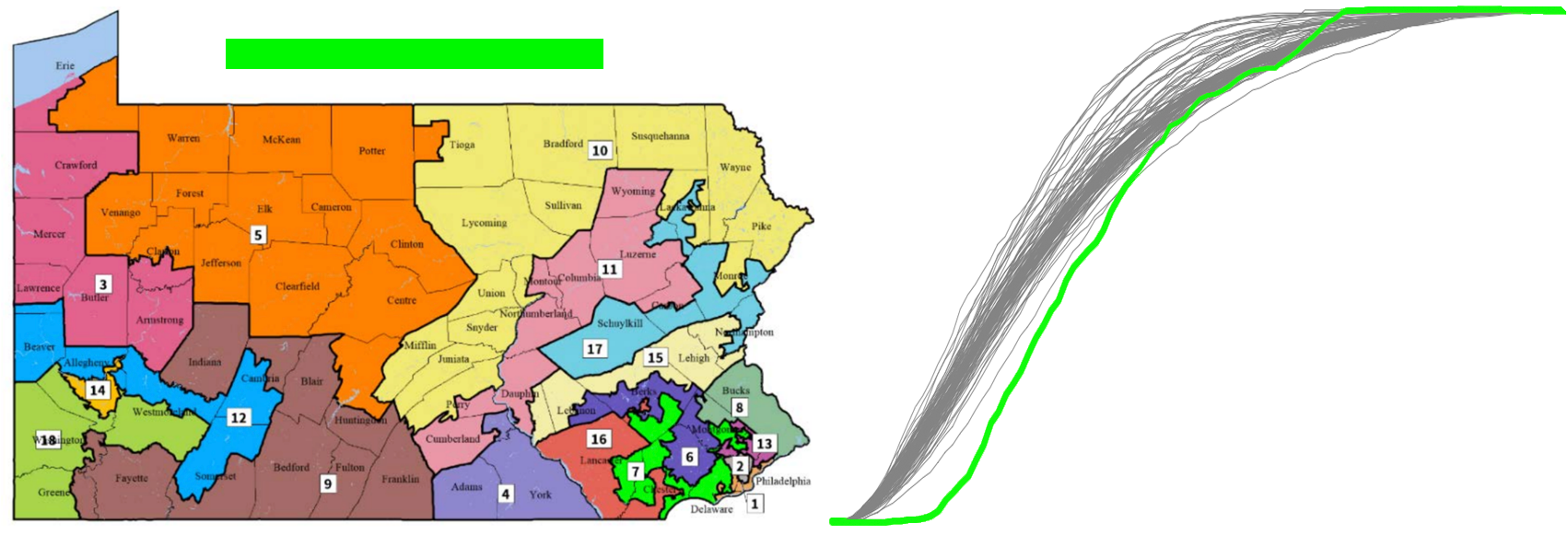}
    \caption{Normalized aDQFs $\tilde{q}_i(\delta)$ based on a kernel defined on persistence diagrams.  Red lines correspond to 99 randomly generated districting maps via ReCom, green corresponds to 2011 map (pictured, source: https://ballotpedia.org/) invalidated by the Supreme Court in 2018 for partisan gerrymandering.}
    \label{fig:gerry_dqf}
\end{figure}

In our analysis, the aDQFs were again computed using the kernel trick via PSSK, resulting in a visualization of the extent at which the 2011 map is anomalous, see Figure \ref{fig:gerry_dqf}.

\section{Implementation}
The computational complexity of the method is somewhat high, due primarily to the construction of a matrix of feature functions, as pairs of points are considered.  Specifically, for each the $O(n^2)$ lines $l_{ij}$, we need to consider the relation of all $n$ points to the base point. We also need to compute the Gram matrix, which in $\R^d$ has order $O(n^2d)$.  Thus, in $\R^d$ the overall complexity is $O(n^3 + n^2d)$. Since we are not working with these individual functions directly but rather their averages, we propose basing these averages on only a subset of these pairwise comparisons, say $\log(n)$ of them. This would reduce the complexity to $O(n^2(\log(n)+d))$.  It is worth noting that the algorithm is also embarrassingly parallel.  

We propose using three visualizations of the (a)DQFs, the averaged (over a random subset) $\bar q_{i}(\delta)$, its normalized (by the maximal value) version,  $\tilde q_{i}(\delta)$, and the first derivative of the normalized version $\frac{d}{d\delta}\tilde q_{i}(\delta)$ (after a small amount of smoothing), at multiple values of $\alpha$ (which as discussed above, does not add much computational complexity to the algorithm).  The functions {\it dqf.outlier} and  {\it dqf.explore} in the R package 
{\it anomalyDQF}\footnote{available at \url{https://github.com/GabeChandler/AnomalyDetection}} generate and visualize these three functions as interactive plots in base-R. 

\begin{figure}[h]
    \centering
    \includegraphics[height=2in, width=4.5in]{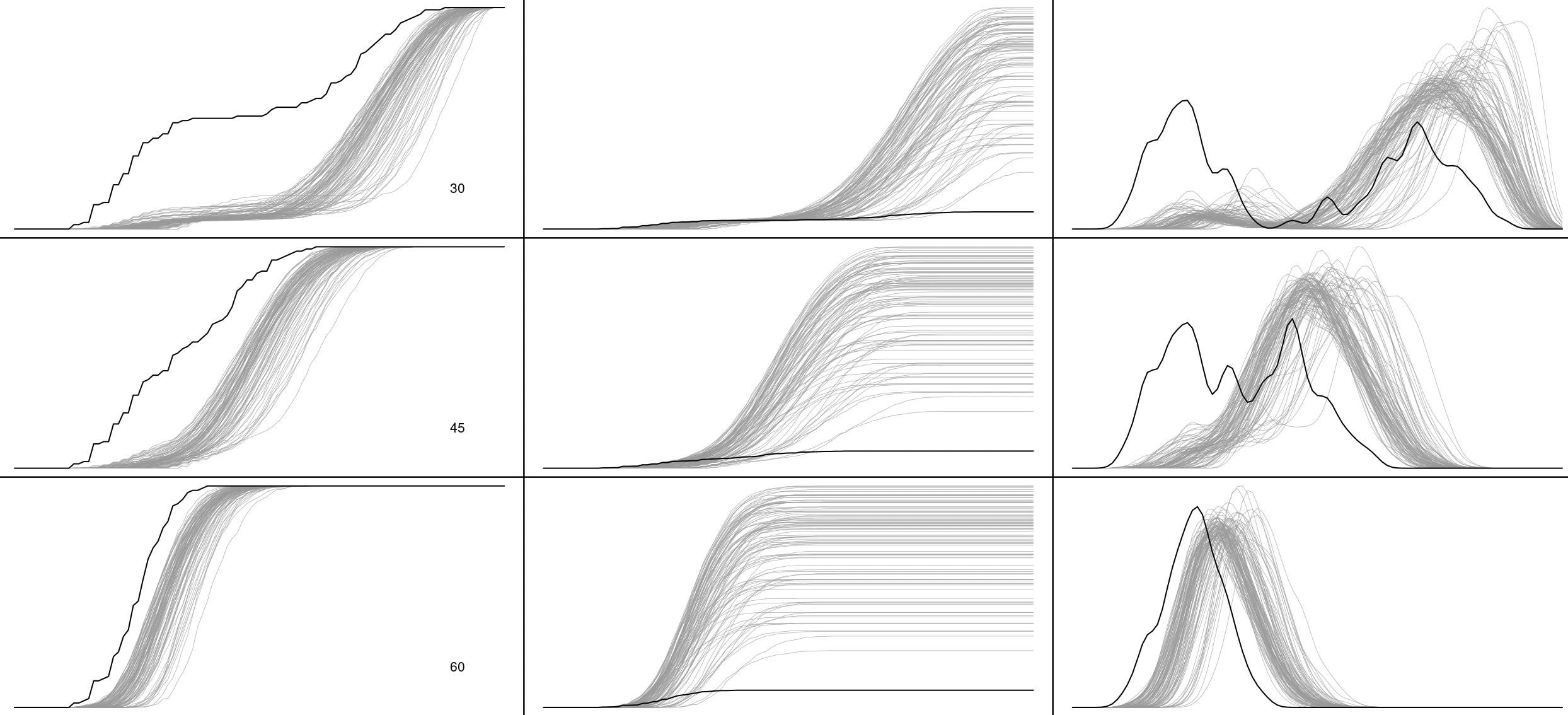}
    \caption{ Visualization of (left column) normalized aDQFs $\tilde{q}_i(\delta)$, (center column) raw averaged aDQFs $\bar{q}_i(\delta)$ and (right column) first derivative of normalized aDQF  $\frac{d}{d\delta}\tilde{q}_i(\delta)$ for the (left panel) $d=30$ simulation data, non-anomalous data living in an annulus, outlier at origin. Outlier indicated by bold.  Rows correspond to different angles ($\alpha=\pi/6, \pi/4, \pi/3$).}
    \label{fig:wine_dqf}
\end{figure}

The 30-dimensional simulation setup used earlier demonstrates the need to consider multiple angles. For a fixed angle and sufficiently high dimension, there is likely to exist a range of $\delta$ such that $\widehat{q}_{ij}(\delta)=\frac{1}{n}$ for all $i,j$ (effectively, a very long flat interval for the average DQF plots for all $i$), as these cones will only include one of the two points $x_i$ and $x_j$ that define the direction.  This is apparent in Figure \ref{fig:wine_dqf} at angle $\alpha=\pi/6$, where we see the pinch point on the middle plot.   

%
Rather than consider the current method as competing with existing outlier detection algorithms, we view the (a)DQF as complementary.  It's clear from the simulation studies and real data examples that both the (a)DQF and other methods are effective at identifying anomalous observations. Many other effective techniques exist as well.  Due to the computational complexity of computing the (a)DQF (mainly with respect to sample size), identifying a suitable subset of the data to visualize via a more computationally friendly method for very large data sets would ease the computational burden.  That is, one can compute the (a)DQF of a random subset of non-anomalous observations and any identified {\it interesting} observations to create a useful visualization of subsets of the data.

\section{Technical proofs.}

{\sc Proof of Lemma~\ref{basic}.} For the heuristic idea of the proof, we refer to the discussion presented right after Lemma~\ref{basic} and the corresponding Figure~\ref{fig:geom-illustr}.

Recall that by definition of $d_x(s)$ (see (\ref{depth-funct})), we have that $\TD(x,F) = \max_s d_x(s),$ and $d_x(s)$ is  $U$-shaped about the minimum value $d_x(x) = 0$, meaning non-increasing for $s \le x$ and non-decreasing for $s > x$. This implies that the sublevel sets $\Gamma(t) := \{s: d_x(s) \le t\}$ are (closed) intervals, while the sets $\Gamma^\circ(t) := \{s: d_x(s) < t\}$ are open intervals.    


%
We split the proof in two parts. First we show that the lemma holds for values $\delta \in \Delta := \big\{\delta_t = G\big(\Gamma^\circ(t)\big),t \in [0,\TD(x,F)]\big\} \subset [0,\delta_x^*]$ (the inclusion will be shown below), and the second part considers $\delta \in [0,\delta_x^*] \setminus \Delta.$ The second part is only needed because $F$ is allowed to have flat parts, for in the special case of $F$ having no flat parts, i.e. the sets  $\{s: d_x(s) = t\}, t \in [0,\TD(x,F)]$ all have Lebesgue measure zero, the map $t \to \delta_t$ is continuous, and consequently $[0,\delta_0^*] \setminus \Delta = \emptyset$ (see below).\\[10pt]
{\sc Claim:} {\em For all $\delta = \delta_t \in \Delta$, we have $q_x(\delta_t) = t$, and $I_{x,\delta_t} := \overline{\Gamma^\circ}(t)$ satisfies the median property (\ref{DQF-rep}).}\\[-10pt]

Note that since $G(I_{x,\delta_t}) = G(\overline{\Gamma^\circ(t)}) = G(\Gamma^\circ(t)) = \delta_t,$ showing the claim means showing the assertion of the lemma for all values of $\delta \in \Delta.$ 

{\sc Proof of claim:} Since $F$ is continuous, $d_x(s)$ is continuous, which follows directly from the definition of $d_x(s)$. This, in turn, implies that the function $[0,\TD(x,F)] \ni t \to {\rm length}(\{s:d_x(s) \le t\})$  is {\em strictly} increasing, and since by assumption $G$ has a strictly positive density, the same holds for $D_x(t) = G(s: d_x(s) \le t),$ the cdf of $d_x(S)$. Note that, by definition, $D_x(t)$ is right-continuous. Next note that the DQF $q_x(\delta)$ is defined as the (left-continuous) generalized inverse of $D_x(t)$, i.e. $q_x(\delta) =  D^-_x(\delta)$. Since $D_x(t)$ is strictly increasing, we know that its generalized inverse $q_x(\delta)$ is indeed continuous. Next, the (left-continuous) generalized inverse of $q_x(\delta)$ is $q^-_x(t) = G(s: d_x(s) < t)$, which is the left-continuous version of $D_x(t).$ Since $q_x(\delta)$ is continuous, we have $q_x(q_x^-(t)) = t$ for all $t$ (a general property of the generalized inverse), and since  $q_x^-(t)= \delta_t\,  (= G(I_{x,\delta_t}))$, this shows the first part of the claim.


To see the second part, note that for any $t \in [0, \TD(x,F)]$, the set $\Gamma^\circ(t)$ is an interval of the form $(a(t),b(t))$, and by continuity of $d_x(s)$, we have that $d_x(a(t)) = d_x(b(t)) = t$. For $F(x) \le 1/2,$ by definition of $d_x(s)$, we have that $d_x(a(t)) = F(a(t)) - F(x)$ and $d_x(b(t)) = F(x) - F(b(t)).$ It follows that $2q_x(\delta_t) = 2t = d_x(a(t)) + d_x(b(t)) = F(b(t)) - F(a(t)) = F(I_{x,\delta_t}).$ This shows that $I_{x,\delta_t}$ satisfies (\ref{DQF-rep}). The case $F(x) > 1/2$ follows similarly. This shows the claim.  \;\;$\Box$

We next show that  it remains to consider the values $[0,\delta^*_x] \setminus \Delta.$ For this, we show that $\Delta \subset [0,\delta_x^*],$ which follows from $\max\{\delta \in \Delta\} = \delta^*_x.$ Since $\max\{\delta \in \Delta\} = G\big(\Gamma^\circ(\TD(x,F))\big) = G\big(\overline{\Gamma^\circ(\TD(x,F))}\big),$ we need to show that $\delta^*_x = G\big(\overline{\Gamma^\circ(\TD(x,F))}\big)$. To see this, we write $\overline{\Gamma^\circ(\TD(x,F))} = [a^*,b^*]$ so that $G(\big(\overline{\Gamma^\circ(\TD(x,F))}\big) = G(b^*) - G(a^*).$ Now, $b^*$ is the smallest value of $s \ge x$ with $d_x(s) = \TD(x,F).$ For $F(x) \le 1/2,$ we have $\TD(x,F) = F(x),$ and thus by definition of $d_x(s)$, $b^*$ is the smallest value $s \ge x$ such that $F(s) - F(x) = F(x),$ so that $b^* = F^-(2F(x)).$ As for the left endpoint, $a^*$, again by definition of $d_x(s)$, $a^*$ is the largest value $s \le x$ such that $F(x) - F(s) = F(x)$, which is the left endpoint $s_*$ of the support of $F$. So we have $\delta_x^* = G(F^-(2F(x))) - G(s_*)$, which is the formula for 
$\delta_x^*$ given in (\ref{delta-star}). The case $F(x) > 1/2$ follows similarly.

It remains to prove the assertion of the lemma for $\delta \in [0,\delta^*_x] \setminus \Delta.$ These are the values $\delta$ that are not in the image of the map $t \to \delta_t$, i.e. they correspond to the (at most countably many) jumps of $t \to \delta_t = G(\overline{\Gamma^\circ(t)})$. Let $[\delta_i,\delta_{i+1}]$ denote such a jump. Then $\delta_i = G(\overline{\Gamma^\circ(t_i)})$ and $\delta_{i+1} = G(\Gamma(t_i))$ for some $t_i \in (0,\TD(x,F)).$ Both $\overline{\Gamma^\circ(t_i)}$ and $\Gamma(t_i)$ are intervals and  $\overline{\Gamma^\circ(t_i)} \subset \Gamma(t_i).$ Since $G$ is continuous, for any $\delta \in [\delta_i,\delta_{i+1}]$, we can find an interval $[a,b]$ with $\overline{\Gamma^\circ(t_i)} \subset [a,b] \subset \Gamma(t_i)$ of $G$-measure $\delta.$ Since further $F(\overline{\Gamma^\circ(t_i)}) = F(\Gamma(t_i))$ (because the flat parts of $d_x(s)$ correspond to flat parts of $F$), the $F$-measure of all these intervals is the same as the $F$-measure of $\overline{\Gamma^\circ(t_i)}$. Moreover, for any such $\delta \in [\delta_i,\delta_{i+1}]$, we also have $q_x(\delta) = q_x(\delta_i)$, because $t \to \delta_t$ is the generalized inverse of $q_x(\delta),$ and thus jumps of $t  \to \delta_t$ correspond to flat parts of $q_x(\delta)$. Thus, property (\ref{DQF-rep}) also holds for the values of $\delta \in [0,\delta_x^*]\setminus\Delta.$ 

To see the last assertion of the lemma, observe that
\begin{align}\label{min-val}\lim_{t\searrow 0} D_x(t) = l_x.
\end{align}
In particular, this implies that for $x \notin S_F$,  $D_x(t)$ has a discontinuity at $t = 0,$ and thus, for such values of $x$, the DQF $q_x(\delta)$ is constant equal to zero for $\delta \in [0,l_x].$ 

\vspace*{-0.8cm}
\begin{flushright}$\Box$\end{flushright}

{\bf Proof of Corollary~\ref{DQF-atzero}.} Observe that we obtain from Lemma~\ref{DQF-rep} that $\frac{2q_x(\delta)}{\delta} = \frac{F(I_{x,\delta})}{G(I_{x,\delta})} = \frac{F(b_{x,\delta}) - F(a_{x,\delta})}{G(b_{x,\delta}) - G(a_{x,\delta})}$, where
$[a_{x,\delta},b_{x,\delta}]$ are the sublevel sets of $d_x(s)$. Let $U$ be an open neighborhood of $x$. For $f(x) \ne 0$ in $U\setminus \{x\}$, the anchor point $x$ is the unique minimizer of $d_x(s)$, and thus both $a_{x,\delta}, b_{x,\delta} \to x$ as $\delta \to 0$ (see also proof of Lemma~\ref{DQF-rep}), and we obtain the result. Now suppose that there exists an open neighborhood $U$ with $f(x) = 0$ for all $x \in U$, this is to say that there exists a `zero-interval' $I_x$ of length $\ell_x > 0$ with $f(x) = 0$ in this interval. As $\delta \to 0$, we have  $[a_{x,\delta},b_{x,\delta}] \to I_x$, and since $F(I_x) = 0$ and $G(I_x) \ne 0$ (since $x \in S_g^\circ$), this implies that $\frac{2q_x(\delta)}{\delta} \to 0$ as $\delta \to 0.$ \begin{flushright}$\Box$\end{flushright}

{\bf Proof of Lemma~\ref{DQF-atzero2}:} By definition of $t^\pm_{x,y}$, we have $F\big(A_{x,y}(s)\big) = 0$ for $s \in \big(m_{x,y} - t^-_{x,y}u,\, m_{x,y} + t^+_{x,y}u) \subseteq  [a_{x,y},b_{x,y}].$ From this we have that $d_{x,y}(s) = \min\big(F(A_{x,y}(s),B_{x,y}(s)\big) = 0.$ The $G_{x,y}$-measure of this set by definition equals $\delta_{x,y}.$ It follows that $q_{x,y}(\delta) = 0$ for $\delta \in [0,\delta_{x,y}],$ and thus have $q_{x,y}(\delta) = 0$ for all $\delta \in [0,  \delta_{x,y}].$ This implies that $\overline q_x(\delta) = 0$ for all $\delta \in [0, \inf_y\delta_{x,y}].$
\begin{flushright}$\Box$\end{flushright}


\acks{The work of W. Polonik was supported in parts by the NSF grant DMS-2015575.}

\section{Appendix: Some background on the HKS and TDA} 
One of our applications is based on the HKS and TDA, and here we provide some more background on these topics.

The construction of the HKS on a compact Riemannian manifold $M$ (without boundary),  such as surface meshes, is as follows. First recall that the heat kernel $h_t(x,y)$ associated to $M$  is the fundamental solution to the heat equation $\big(\Delta_x - \frac{\partial}{\partial t}\big)u(x,y,t) = 0$ with domain $M$ and initial condition $u(x,y,0) = \delta_x(y)$, where $\Delta_x(u) = {\rm div}\,{\rm grad}(u)$ denotes the Laplace-Beltrami operator associated to $M$. For instance, for $ M = \R^d$ the heat kernel is $h_t(x,y) = \frac{1}{(4\pi t)^{d/2}}e^{-\frac{\|x-y\|^2}{4t}}.$ The heat kernel on $M$ admits an eigen-decomposition of the form $h_t(x,y) = \sum_{j=1}^\infty e^{-\lambda_j t} \varphi_j(x)\varphi_j(y)$ with $(\lambda_j,\varphi_j)$ being eigenpairs of the Laplace-Beltrami operator. The HKS $\tilde h_t(x)$ then is the diagonal of the heat kernel, i.e. $\tilde h_t(x) = h_t(x,x) = \sum_{j=1}^\infty e^{-\lambda_j t} \varphi^2_j(x).$ As a function of the eigenvalues and eigenvectors of the estimated Laplace-Beltrami operator, the HKS  carries information on geometric properties of the manifold (see Sun et al., 2009). Indeed, under mild assumptions the HKS, if considered as a function of both $t$ and $x$, characterizes the manifold up to isometries, and it thus lends itself to shape classification. Moreover, we do not use the HKS as a function in both $t$ and $x$, but we pick a fixed value of $t$ (a tuning parameter). For `small' values of $t$, the HKS contains curvature information. Indeed, it is known that $\tilde h_t(x) = \big(\frac{1}{4\pi t}\big)^{d/2} (1 + \frac{1}{6}s(x)t + O(t^2))$, where $s(x)$ is the Gaussian curvature at $x$ and $d$ is the dimension of $M$.

In our application, the shapes come as meshes, and since the heat kernel (and thus the HKS) is only known explicitly for some exceptional cases, we need to approximate the HKS for the given shapes. The approximation of the HKS that is being used is based on the graph Laplacian on the mesh. For details on graph Laplacians on meshes, see e.g. Belkin et al. (2008).  It is a  piece-wise linear function $\hat h_t(x)$ on the surface mesh with a constant value on each face of the mesh. This piece-wise linear function is constructed by first defining function values $\hat h_t(v_i)$ on all the vertices $v_i,i=1,\ldots,N$ (where $t$ is a tuning parameter), and then defining the function value on each face $f_j$ of the mesh as $\hat h_t(f_j)= \max_{v_i \in f_j}\hat h_t(v_i)$. The values $\hat h_t(v_i)$ have the form $\hat h_t(v_i) = \sum_{k=1}^N e^{-\lambda_k t}\phi^2_{ik},$ where $\phi_i = (\phi_{i1},\ldots,\phi_{iN})$ is the $i^{th}$ eigenvector of a normalized graph Laplacian with $\lambda_i$ the corresponding eigenvalue. Also note that here we consider the HKS as a function in $x$, while $t$ is a tuning parameter. Comparing HKSs for different meshes directly is not possible, because they live on different surfaces. Thus, certain features of these functions are extracted. This is accomplished by constructing a persistence diagram for each of the HKSs, using their (lower) level set filtration. We refer to Chazal and Michel (2021) or Wasserman (2018) for introductions to persistent homology, including persistence diagrams. In a nutshell, these persistence diagrams measure the dynamics of the number of topological features of the level sets of the HKSs as the level increases along with their persistence (significance). By topological features we mean connected components (in this context also called $0$-dimensional holes), and $k$-dimensional holes, with $k \ge 1$, where in practice $k \le 2.$ Heuristically, a $k$-dimensional hole is enclosed by a $k$-dimensional surface. More formally, topological features correspond to basis vectors of homology groups (vector spaces). The persistence diagram generated by the sublevel set filtration of the HKS then arises by considering pairings of critical points of the HKS (for more details on how these pairings are formed see for instance Chazal and Michel, 2021). The smaller of the two heights of two paired critical points is called the birth time of a topological feature and the larger is its death time. The persistence diagram then is a 2D plot of the birth times against the death times of all the topological features. Persistence of a feature is the difference between birth and death times. It is these pairings that gives rise a global measure of topology/geometry out of the local properties of being critical points. A small persistence is interpreted as the critical point being a `small bump', and might therefore be generated by noise. Large persistences, on the other hand, are interpreted as (topological) signal.  Comparing such 2-dimensional point sets can be done by using the kernel trick to implicitly map the persistence diagram into an RKHS and to then use the corresponding RKHS geometry in our DQF approach. The kernel that exists for this purpose is the persistence scale-space kernel (PSSK) $k_{\sigma}$ of Reininghaus et al. (2015) given by
$$ k_\sigma({\rm Dgm}_1, {\rm Dgm}_2) = \frac{1}{8\pi\sigma} \sum_{p \in {\rm Dgm}_1\atop q \in {\rm Dgm}_2} e^{-\frac{1}{8\sigma^2}\|p - q\|^2} - e^{-\frac{1}{8\sigma^2}\|p - \overline{q}\|^2},$$
where ${\rm Dgm}_1, {\rm Dgm}_2$ denote two persistence diagrams, and $\overline q$ denotes the reflection on the diagonal of $q.$ This kernel is constructed by solving the heat diffusion on the persistence diagram, meaning that the initial conditions in the heat equation are point masses (Dirac deltas) in the point of the persistence diagram, and enforcing the boundary condition of being zero on the diagonal so that points close to the boundary, i.e. points with small persistence, contribute little, reinforcing the heuristic interpretation of persistence given above.

\section{References}
Aggarwal, C.C. (2013): Outlier analysis. Springer, New York.\\[5pt]

Belkin, M., Sun, J. and Wang, Y. (2008): Discrete Laplace operator on meshed surfaces. In {\it SCG '08: Proceedings of the twenty-fourth annual symposium on computational geometry}, 278 - 287.\\[5pt]
Bishop, C.M. (1993). Novelty Detection and Neural Network Validation. In: Gielen, S., Kappen, B. (eds), ICANN 1993. Springer, London.\\[5pt]
Blake C.L., and Merz C.J. (1998): UCI Repository of Machine
  Learning Databases. University of California, Irvine,
  Department of Information and Computer Sciences, Irvine,
  CA.\\[5pt]
Bouman, R., and Heskes, T. (2025): Autoencoders for Anomaly Detection are Unreliable. 10.48550/arXiv.2501.13864.\\[5pt]
Burridge, P., and Taylor, R. (2006): Additive outlier detection via extreme-value theory. {\em J. Time Ser. Anal}. {\bf 27}. 685-701.\\[5pt]
Chandler, G. and Polonik, W. (2021): Multiscale geometric feature extraction for high-dimensional and non-Euclidean data with applications. {\em Ann. Statist.} {\bf 49}, 988-1010.\\[5pt]
Chandola, V., Banerjee, A., and Kumar, V. (2009). Anomaly detection: A
survey. ACM Computing Surveys, 41(3):15, 1–58.\\[5pt]
Chazal, F. and  Michel, B. (2021): An introduction to topological data analysis: fundamental and practical aspects for data scientists. {\em Front. Artif. Intell.}, 4.\\[5pt]
Chen, Y., Hao, Y., Rakthanmanon, T., Zakaria, J., Hu, B., and Keogh, E. (2015): A general framework for never-ending learning from time series streams. {\em Data Min. Knowl. Discov.}. 29. \\[5pt]
Cortes, D. (2021): Revisiting randomized choices in isolation forests. 10.48550/arXiv.2110.13402. \\[5pt]
Cortes, D. (2025): isotree: Isolation-Based Outlier Detection. R
  package version 0.6.1-4,
  \url{https://CRAN.R-project.org/package=isotree}.\\[5pt]
DeFord, D., Duchin, M. and Solomon, J. (2021): Recombination: A family of Markov chains for redistricting. {\em Harvard Data Sci. Rev.}  3.1
\\[5pt]
Dubey, P., Chen, Y., and M{\"u}ller, H.G. (2024). Metric statistics: Exploration and inference for random objects with distance profiles. {\it Ann. Statist.}, 52(2), 757-792.\\[5pt]
Duchin, M., and Needham, T. and Weighill, T. (2021): The (homological) persistence of gerrymandering. {\em Found. Data Sci.} 10.3934/fods.2021007. \\[5pt]
Einmahl, J.H.J., Gantner, M., and Sawitzki, G. (2010a): The shorth plot. {\em J. Comput. Graphical Stat.}, {\bf 19}(1), 62-73.\\[5pt]
Einmahl, J.H.J., Gantner, M., and Sawitzki, G. (2010b): Asymptotics of the shorth plot. {\em J. Stat. Plann. Inference}, {\bf 140}, 3003–3012.\\[5pt]
Fefferman, C., Mitter, S., and Narayanan, H. (2016): Testing the manifold hypothesis. {\em J. Am. Math. Soc.} {\bf 29}, 983–1049.\\[5pt]
Fisherkeller, M.A., Friedman, J.H. and Tukey, J.W. (1974): PRIM-9, an interactive multidimensional data display and analysis system. {\it Proceedings of the Pacific ACM Regional Conference}. [Also in The Collected Works of John W. Tukey V (1988) pp. 307–327.]\\[5pt]
Friedman, J.H. and Stuetzle, W. (2002): John W. Tukey's work on interactive graphics. {\em Ann. Stat.}, {\bf 30}, 1629-1639.\\[5pt]
Hawkins, D.M. (1980): Identification of outliers. Chapman and Hall, London – New York.\\[5pt]
Huber, P.J. (1964): Robust estimation of a location parameter. {\it Ann. Math. Stat.},
{\bf 35}, 73–101.\\[5pt]
Hyndman, R.J. (1996): Computing and graphing highest density regions. {\em Am. Stat.}, {\bf 50}, 120-126.\\[5pt]
Hodge, V.J. and Austin, J. (2004): A survey of outlier detection methodologies. {\em Artif. Intell. Rev.}, {\bf 22} (2), 85-126. \\[5pt]
Lientz, B.P. (1974): Results on nonparametric modal intervals, {\em SIAM J. Appl. Math}, {\bf 19}, 356-366.\\[5pt]
Liu, F.T., Ting, K.M., and Zhou, Z. (2008): Isolation forest. {\em 2008 Eighth IEEE International Conference on Data Mining},  413–422.\\[5pt]
Meyer D., Dimitriadou, E., Hornik, K., Weingessel, A., and Leisch, F. (2022): e1071: Misc Functions of the Department of Statistics, Probability Theory Group (Formerly:
  E1071), TU Wien. R package version 1.7-11,
  \url{https://CRAN.R-project.org/package=e1071}.\\[5pt]
Mozharovskyi, P., and Valla, R. (2025): Anomaly detection using data depth: multivariate case. {\it Int. J. Data Sci. Anal.} \\[5pt]
Minnotte, M. and Scott, D. (1993): The Mode Tree: A tool for visualization of nonparametric density features. {\it J. Comput. Graphical Stat.} {\bf 2}, 51-68.\\[5pt]
Nagy, S., Gijbels I., and Hlubinka, D. (2017): Depth-Based recognition of shape outlying functions, {\em J. Comput. Graphical Stat.}, {\bf 26}, 883-893.\\[5pt]
%
%
Ojo, O., Lillo, R., and Fernandez Anta, A. (2023): fdaoutlier: Outlier Detection Tools for
  Functional Data Analysis. R package version 0.2.1,
\url{https://github.com/otsegun/fdaoutlier}.\\[5pt]
Reininghaus, J., Huber, S., Bauer, U. and Kwitt, R. (2015): A stable multi-scale
kernel for topological machine learning. {\em IEEE Conference on Computer Vision and
Pattern Recognition (CVPR)}, 4741–4748.\\[5pt]
R\'{e}nyi, A. and Sulanke, R. (1963):  Über die konvexe Hülle von n zufällig gerwähten Punkten {\em Z. Wahrscheinlichkeitstheor. Verw. Geb.}, {\bf 2}, 75–84.\\[5pt]
Ruff, L., Kauffmann, J., Vandermeulen, R., Montavon, G.,  Samek, W.,   Kloft, M., Dietterich, T., and Müller, K. (2021): A Unifying Review of Deep and Shallow Anomaly Detection. {\em Proceedings of the IEEE.} {\bf 109}(5), 1-40. \\[5pt]
Sun, J., Ovsjanikov, M. and Guibas, L. (2009): A concise and probably informative multi-scale signature based on heat diffusion. {\em Computer Graphics Forum}, {\bf 28}(5), 1383 - 1392.\\[5pt]
Sawitzki, G. (1994): Diagnostic plots for one-dimensional data. In: Ostermann, R., Dirschedl, P. (Eds.), {\em Computational Statistics, 25th Conference on
Statistical Computing at Schloss Reisensburg.} Physica-Verlag, Springer, Heidelberg, pp. 237–258.\\[5pt]
B. Sch\"{o}lkopf, J.C. Platt, J., Shawe-Taylor, A.J. Smola and R.C. Williamson (2001): Estimating the support of a high-dimensional distribution, {\em Neural Computation,} {\bf  13}(7), 1443-1471.\\[5pt]
Talagala, P.D. (2020): stray: Anomaly Detection in High Dimensional and Temporal Data.
  R package version 0.1.1, \url{https://CRAN.R-project.org/package=stray}.\\[5pt]
Talagala, P.D., Hyndman, R.J. and Smith-Miles, K. (2021): Anomaly detection in high-dimensional data, {\em J. Comput. Graphical Stat.}, DOI: 10.1080/10618600.2020.1807997\\[5pt]
Tenenbaum J.B., de Silva V., and Langford J.C. (2000): A global geometric framework for nonlinear dimensionality reduction. {\em Science}. {\bf 290}(5500), 2319-23.\\[5pt]
Tukey, J.W. (1975): Mathematics and the picturing of data. {\it Proceedings of the International Congress of Mathematicians. 1975}, pp. 523–531.\\[5pt]
Wasserman, L. (2018): Topological data analysis. {\em Annu. Rev. Stat. Appl.}, {\bf 5}, 1501-532.\\[5pt]
Wilkinson, L. (2017): Visualizing big data outliers through distributed aggregation. {\em IEEE Trans. Visual Comput. Graphics} {\bf 24}(1), 256–266.
\end{document}